\documentclass[twocolumn,showpacs,prb,citeautoscript,amsmath,amssymb,floatfix,eqsecnum]{revtex4-2}

\usepackage{amsmath,amssymb,color}
\usepackage{bm}
\usepackage{graphicx}
\usepackage{psfrag}
\usepackage{bbold}
\usepackage{bbm}
\usepackage{hyperref}
\newcommand{\bd}{\bm}
\newcommand{\kv}{{\bm{k}}}

\begin{document}

\title{
Functional renormalization group in an external field:   \\
constant field versus constant order parameter}

\author{Julia von Rothkirch, 
Andreas R\"{u}ckriegel, and 
Peter Kopietz}
  
\affiliation{Institut f\"{u}r Theoretische Physik, Universit\"{a}t
  Frankfurt,  Max-von-Laue Strasse 1, 60438 Frankfurt, Germany}

\date{October 2, 2026}

 \begin{abstract}
Renormalization group (RG) flows for systems subject to an external  field can be derived  either by fixing the field or by fixing the conjugate order parameter. We show how to implement these boundary conditions within the
functional renormalization group approach.
We derive and compare the corresponding flow equations for $\phi^4$ theory 
and
show that some scaling variables and scaling exponents
associated with the linearized flow of the relevant three-point vertex 
in the vicinity of the Ising fixed point depend on 
the boundary conditions.
We also discuss the effect of elasticity on Ising criticality. In this case two of the fixed points (namely the renormalized Ising fixed point and the spherical fixed point) obtained for constant strain are shifted to  physically inaccessible regimes of parameter space  when we perform the RG at constant stress. 

\end{abstract}


\maketitle

\section{Introduction}

Since its invention in the 1970s 
the renormalization group (RG) has given deep insights into
the physics of phase transitions and critical phenomena \cite{Wilson72,Wilson74,Ma76,Fisher83,Goldenfeld92,Fisher98}. In particular, Ising criticality close to 
$D=4$ dimensions can be understood by applying  RG methods to the Ginzburg-Landau-Wilson 
model with bare action
\begin{equation}
 S [ \phi; h_0 ] = \int d^D r \left[ 
 \frac{c_0}{2} ( {\mathbf{\nabla}} \phi )^2  + \frac{r_0}{2} \phi^2
 + \frac{u_0}{4!} \phi^4 - h_0 \phi \right],
 \label{eq:Sdef}
 \end{equation}
where the scalar field $\phi ( \bd{r} ) $  represents the fluctuations of the magnetization while  $h_0$  is an external magnetic field.
RG transformations map the bare action (\ref{eq:Sdef}) onto an effective  coarse grained action depending on a scale parameter $\Lambda$. In the effective action  
the bare couplings  $c_0$, $r_0$, $u_0$, and $h_0$  are replaced by scale-dependent couplings  $c_\Lambda$, $r_\Lambda$, $u_\Lambda$, and
$h_\Lambda$.
In addition, the RG generates infinitely many new couplings 
most of which can be classified as irrelevant. 
For vanishing magnetic field  $h_0=0$ the
RG flow equations describing the evolution of the couplings
$c_\Lambda$, $r_\Lambda$ and $u_\Lambda$ with the scale $\Lambda$ have been derived a
long time ago by the 
pioneers of the renormalization group \cite{Wilson72,Wilson74,Ma76}. Close to four dimensions the smallness 
of $\varepsilon = 4-D$ can be used to calculate the fixed points of the RG transformation and the associated critical exponents in a controlled way.
Detailed descriptions of this calculation using either the elementary momentum shell technique or the more sophisticated
functional renormalization group (FRG) can be found in textbooks \cite{Ma76,Kopietz10}. 
However, at some early point in the derivation of the RG flow the external magnetic field is usually  set equal to zero.  Surprisingly, we have not been able to find in the literature a comprehensive derivation of the RG flow equations for the model (\ref{eq:Sdef}) including the magnetic field and
the  relevant three-point vertex induced by it. 
For imaginary magnetic fields the results of a perturbative RG calculation  close to six dimensions have been reported by Fisher \cite{Fisher78}, who pointed out that in this case the
Yang-Lee edge singularity \cite{Yang52,Lee52} in the distribution of zeros of the partition function is closely analogous to an ordinary critical point that can be described by an effective $\phi^3$ theory.
An obvious question arising in this context is whether the magnetic field or the conjugate magnetization $m = \langle \phi \rangle$ should be fixed  during the RG transformation.
In fact, it is a priori not clear whether the  RG flow at constant magnetic field
is characterized  by the  same fixed points as the RG flow at constant magnetization. 
In this work we will use a FRG approach to examine how this change in external conditions
affects the RG flow.

To put this problem into a broader context, we will also study a more general field theory where
an Ising order parameter is coupled to the volumetric strain of an underlying lattice.
In this case the stress  conjugate to the volumetric strain can be identified with the pressure.
The RG transformation can be performed either for constant strain or for constant pressure.  In the former case, the RG exhibits four distinct fixed points \cite{Bergman76,Hansen25}: apart from the Gaussian fixed point  G and the Ising fixed point I, one obtains two additional fixed points 
S (spherical) and R (renormalized Ising) whose critical exponents can be obtained from the critical exponents associated with  G and I by Fisher renormalization \cite{Fisher68}. 
On the other hand, 
we show in this work that
when the RG is performed at  constant pressure the fixed points S and R are shifted to unphysical regimes of parameter space so that only G and I are accessible by properly fine-tuning the initial conditions.

The rest of this work is organized as follows:
We begin in Sec.~\ref{sec:RGm} with the derivation of the RG flow of the
model (\ref{eq:Sdef}) for fixed magnetization $m = \langle \phi \rangle$ and small
$\varepsilon = 4 -D$.  We show how the boundary condition of constant $m$ implies an exact flow equation for the scale-dependent  external magnetic field $h_{\Lambda}$. Moreover, we 
explicitly calculate the relevant scaling fields for the Wilson-Fisher fixed point and show how to obtain the critical magnetic equation of state.
In Sec.~\ref{sec:RGh} we repeat the calculation with the alternative boundary condition that the external magnetic field $h= h_0$ is fixed during the RG flow, implying that the magnetization $m_{\Lambda}$ is scale-dependent. 
While the existence of the Ising fixed point is found to be independent of the boundary conditions, the exponent controlling the scaling of the three-point vertex changes
when we fix $h$ instead of $m$. 
As an example for a model where a change in boundary conditions can 
actually change the number of physically accessible fixed points,
we consider
in Sec.~\ref{sec:RGel} a critical Ising system 
described by the $\phi^4$ theory given in Eq.~(\ref{eq:Sdef}) 
that is coupled to volumetric strain. 
Finally, in Sec.~\ref{sec:summary} we summarize our results and present our conclusions.

\section{FRG for constant  magnetization}
\label{sec:RGm}

\subsection{General procedure}

To set up our notation, let us briefly recall the main steps in the derivation of the
Wetterich equation for our model \cite{Wetterich93,Berges02,Kopietz10,Pawlowski07,Dupuis21}. 
Transforming the order-parameter field to momentum space,
$\phi_{\bd{k}} = \int d^D r e^{ - i \bd{k} \cdot \bd{r} } \phi ( \bd{r} )$, 
our bare action (\ref{eq:Sdef}) can be written as
\begin{align}
&  S [ \phi ; h_0 ] 
 =    - h_0 \phi_{\bd{k} =0} +  \frac{1}{2} \int_{\bd{k}}  \left( c_0 k^2 + r_0  \right) \phi_{- \bd{k}} \phi_{\bd{k}} 
\nonumber
 \\
  &  + \frac{u_0}{4!} \int_{\bd{k}_1}   \ldots  \int_{\bd{k}_4}
 {\cal{V}} \delta_{ \bd{k}_1 + \bd{k}_2 + \bd{k}_3 + \bd{k}_4 ,0 } \phi_{\bd{k}_1 } \phi_{\bd{k}_2} \phi_{\bd{k}_3}
 \phi_{\bd{k}_4},
 \label{eq:Smom}
 \end{align}
where ${\cal{V}}$ is the volume of the system and
$\int_{\bd{k}} = {\cal{V}}^{-1} \sum_{\bd{k}}$.
Let us now  modify the bare action (\ref{eq:Sdef}) by introducing a regulator $R_{\Lambda}$ depending on a momentum scale $\Lambda$,
\begin{equation}
S_{\Lambda} [ \phi ; h_0 ] = S [ \phi ;  h_0 ] +  \frac{1}{2}
( \phi, R_{\Lambda} \phi ),
 \label{eq:Sreg}
 \end{equation}
where in momentum space the Gaussian regulator term can  be written as
 \begin{equation}
  \frac{1}{2} ( \phi, R_{\Lambda} \phi ) = \frac{1}{2}
 \int_{\bd{k}}   \phi_{- \bd{k}}   R_{\Lambda}( \bd{k} )   \phi_{\bd{k}} .
 \end{equation}
The regulator $R_{\Lambda} ( \bd{k} )$ should vanish
for $\Lambda \rightarrow 0$ so that
we recover our original model, while  
$\lim_{\Lambda \rightarrow \infty} R_{\Lambda} ( \bd{k} ) = \infty$ 
so that fluctuations 
on all scales are suppressed for $\Lambda \rightarrow \infty$.
The momentum scale $\Lambda$ separates short-wavelength from long-wavelength fluctuations.
Given the regulated action, 
we define the generating functional of connected correlation functions via
the functional integral
 \begin{align}
 e^{ {\cal{G}}_{\Lambda} [ J  ] } & = 
 \int {\cal{D}}[  \phi ] e^{ - S_{\Lambda} [ \phi ; h_0=0] + 
 ( J, \phi ) }
 \nonumber
  \\
  & = \int {\cal{D}} [ \phi ]
 e^{ - S [ \phi ; h_0=0  ] - \frac{1}{2} ( \phi, R_{\Lambda}  \phi ) + ( J , \phi ) },
 \label{eq:Ghdef}
 \end{align}
where
 \begin{align}
  ( J , \phi ) &  
  = \int d^D r J ( \bd{r} ) \phi ( \bd{r} )
  = \int_{ \bd{k}} J_{ - \bd{k}}  \phi_{\bd{k}}
  .
 \end{align}
Note that the regulated action $S_{\Lambda} [ \phi ; h_0=0]$ 
in Eq.~(\ref{eq:Ghdef}) is taken for vanishing magnetic field, 
because $h_0$ can be absorbed into the uniform part of the source field $J ( \bd{r} )$ and
all connected correlation functions for finite field can be generated by setting the source to $J ( \bd{r} )=h_0$ after taking appropriate functional derivatives.
The scale-dependent 
average effective action 
is now defined as usual~\cite{Kopietz10,Berges02,Wetterich93} via the following subtracted Legendre transform:
 \begin{align}
 \Gamma_{\Lambda } [ \bar{\phi}  ] & =   ( J , \bar{\phi} )
 - {\cal{G}}_{\Lambda} [ J  ]
 - \frac{1}{2} ( \bar{\phi} , R_{\Lambda} \bar{\phi} ).
 \label{eq:Gammahdef}
 \end{align}
On the right-hand side we should express the source 
$J = J [ \bar{\phi}  ]$ as 
functional of the field average $\bar{\phi}$  by inverting the relation
 \begin{equation}
 \frac{ \delta {\cal{G}}_{\Lambda} [ J ] }{ \delta J_{ - \bd{k}} }
  = \langle \phi_{\bd{k}} \rangle_J  \equiv \bar{\phi}_{\bd{k}} ,
 \label{eq:GJm}
 \end{equation}
where the average $\langle \phi_{\bd{k}} \rangle_J$ is taken for fixed sources $J ( \bd{r} )$.
By construction the first functional derivative of 
$\Gamma_{\Lambda} [ \bar{\phi}   ]$ 
is then
 \begin{align}
  \frac{ \delta \Gamma_{\Lambda} [ \bar{\phi}   ]}{\delta \bar{\phi}_{ - \bd{k} } } =  J_{\bd{k} } - R_{\Lambda} ( \bd{k} ) \bar{\phi}_{  \bd{k}} .
 \label{eq:Gamma1JR}
 \end{align}
The RG flow of the functional $\Gamma_{\Lambda} [ \bar{\phi} ]$ is determined by the the Wetterich equation  \cite{Kopietz10,Berges02,Wetterich93}  
 \begin{equation}
 \partial_{\Lambda} \Gamma_{\Lambda} [ \bar{\phi}   ] = 
 \frac{1}{2} {\rm Tr} \left[ ( \mathbf{\Gamma}^{\prime \prime}_{\Lambda} [ \bar{\phi} ] + \mathbf{R}_{\Lambda} )^{-1} \partial_\Lambda \mathbf{R}_{\Lambda} \right],
  \label{eq:Wetterich}
 \end{equation}
where the trace is over momentum space and
the bold symbols
$ \mathbf{\Gamma}^{\prime \prime}_{\Lambda} [ \bar{\phi}  ]$ and 
$\mathbf{R}_{\Lambda}$ are matrices in momentum space 
with matrix elements given by
 \begin{align}
 \bigl[  \mathbf{\Gamma}^{\prime \prime}_{\Lambda} [ \bar{\phi}  ] \bigr]_{ \bd{k} 
 \bd{k}^{\prime}}
  & =     \frac{ \delta \Gamma_{\Lambda} [ \bar{\phi}  ] }{ \delta \bar{\phi}_{\bd{k}}
 \delta \bar{\phi}_{\bd{k}^{\prime}}},
 \label{eq:funcdef}
 \end{align}
and
 \begin{align}
 \bigl[ \mathbf{R}_{\Lambda} \bigr]_{ \bd{k}  \bd{k}^{\prime}} &  =  
 {\cal{V}} \delta_{\bd{k} + \bd{k}^{\prime} ,0  }  
R_{\Lambda} ( \bd{k} ) .
 \end{align}
Expanding both sides of the Wetterich equation (\ref{eq:Wetterich}) in powers of the field yields an infinite hierarchy of exact flow equations for the regularized irreducible vertices.

In this section we impose the boundary condition that the magnetization
$\langle \phi ( \bd{r} ) \rangle_{h_\Lambda}  =m$ is a constant  independent of the RG scale $\Lambda$, where the average
$\langle \phi ( \bd{r} ) \rangle_{h_\Lambda} $ is taken for a homogeneous source $J( \bd{r}) = h_{\Lambda}$. 
The fact that for constant $m$ the magnetic field $h_{\Lambda}$ must be scale dependent  follows from Eq.~(\ref{eq:GJm}), which implies in real space
\begin{equation}
 \left. \frac{ \delta {\cal{G}}_{\Lambda} [ J ] }{\delta J ( \bd{r})  }
  \right|_{ J  ( \bd{r} ) =h_\Lambda} = m.
\label{eq:Ghm1}
 \end{equation}
Since the right-hand side is independent of $\Lambda$, the implicit
$\Lambda$-dependence due to $h_{\Lambda}$ must  cancel the explicit $\Lambda$-dependence due to ${\cal{G}}_{\Lambda} [ J ]$. 
To derive the equivalent boundary condition for
the generating functional $\Gamma_{\Lambda} [ \bar{\phi} ]$,
it is convenient to choose the regulator $R_{\Lambda} ( \bd{k} )$ such that $R_{\Lambda} ( \bd{k} =0 ) =0$, 
which can be implemented by setting 
 \begin{equation}
R_{\Lambda} ( \bd{k} ) = (1 - \delta_{\bd{k} , 0 } ) R ( k^2 / \Lambda^2 ),
 \label{eq:Rdef}
 \end{equation}
  where $R (x )$ is some dimensionless function satisfying
$R ( \infty ) =0$ and $R (0) =1$. 
This amounts to regulating only fluctuations with finite momentum $\bd{k} \neq 0$,  while
homogeneous fluctuations  with momentum $\bd{k} =0$ are  not affected by the regulator.
The reason why it is neither necessary nor convenient to regulate the homogeneous fluctuations with momentum $\bd{k} =0$ is that 
the effective action of the homogeneous mode is proportional to the volume of the system, so that 
in the thermodynamic limit fluctuations of this mode can be neglected and
we may use the saddle-point approximation to calculate the
contribution of the homogeneous mode to the free energy \cite{Schuetz06,Sinner08,Kopietz10}.
The regulator-dependent last term in Eq.~(\ref{eq:Gamma1JR}) then vanishes for homogeneous fluctuations where $\bar{\phi}_{\bd{k}} =
{\cal{V}} \delta_{\bd{k} =0}  m$,
so that Eq.~(\ref{eq:Gamma1JR}) implies in real space
\begin{equation}
 \left. \frac{ \delta {\Gamma}_{\Lambda} [ \bar{\phi} ] }{\delta 
 \bar{\phi}  ( \bd{r})  }
  \right|_{ \bar{\phi} ( \bd{r} ) =m}
  =  h_{\Lambda},
  \label{eq:Gammamh1}
 \end{equation}
and in momentum space~\cite{footnotefunc}
  \begin{equation}
  \left. \frac{ \delta \Gamma_{\Lambda} [ \bar{\phi} ]}{\delta      \bar{\phi}_{\bd{k}}}
  \right|_{  \bar{\phi}_{\bd{k}}  
   = {\cal{V}} \delta_{\bd{k},0} m}  
 = {\cal{V}} \delta_{\bd{k},0} h_{\Lambda}  .
  \label{eq:hlR}
 \end{equation}
At this point it is convenient to shift $\bar{\phi} ( \bd{r} ) = m + \varphi (\bd{r} )$ and define  vertices for fixed magnetization via the functional Taylor  expansion  of the shifted functional $\Gamma_{\Lambda} [ m + \varphi ]$, 
\begin{align}
 {\Gamma}_{\Lambda} [ m + \varphi ]   = {\Gamma}_{\Lambda}^{(0)}
  + &  \sum_{ n =1}^{\infty} \frac{1}{n!} \int_{\bd{k}_1} \ldots \int_{\bd{k}_n }
  {\cal{V}} \delta_{ \bd{k}_1 + \ldots + \bd{k}_n , 0 }
  \nonumber
   \\
   & \times 
  {\Gamma}^{(n)}_{\Lambda} ( \bd{k}_1 , \ldots , \bd{k}_n )
  \varphi_{ \bd{k}_1 } \ldots \varphi_{\bd{k}_n } ,
  \label{eq:Gammashiftexp}
\end{align}
where all vertices ${\Gamma}^{(n)}_{\Lambda} ( \bd{k}_1 , \ldots , \bd{k}_n )$ depend implicitly on the magnetization $m$.
The boundary condition (\ref{eq:hlR}) implies that
the one-point vertex can be identified with the renormalized magnetic field,
 \begin{equation}
 \Gamma^{(1)}_{\Lambda} ( \bd{k}=0 ) = h_{\Lambda}.
 \end{equation}

\subsection{Flow equations for constant $m$}

\begin{figure}
 \begin{center}
  \centering
 \includegraphics[width=\columnwidth]{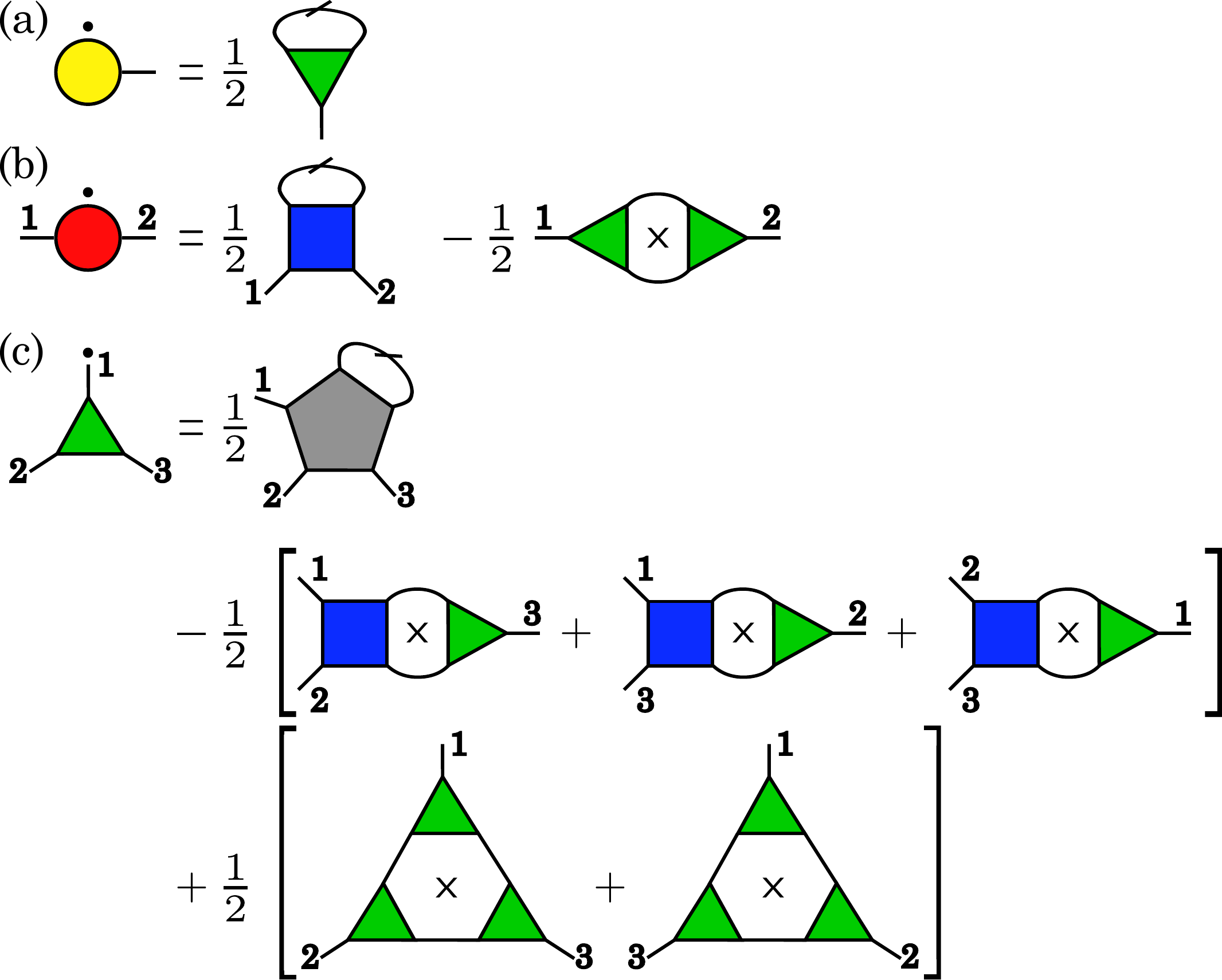}
   \end{center}
  \caption{%
Graphical representations of exact FRG flow equations for constant order parameter $m$:
(a) Flow of the one-point vertex, which is represented by a yellow bubble with one external leg; see  Eq.~(\ref{eq:hflow}). Note that
according to Eq.~(\ref{eq:Gammamh1}) the one-point vertex can be identified
with the scale-dependent magnetic field $h_{\Lambda}$. (b) Flow of the two-point vertex represented by a red  bubble with two external legs. (c) Flow of the three-point vertex represented by a green triangle with three external legs.  
The grey pentagon denotes the five-point vertex and
the dot above the vertices on the left-hand side represents the scale-derivative
$\partial_{\Lambda}$. Moreover, solid lines represent the propagator $G_\Lambda ( \bd{q} )$ while
slashed solid lines represent the corresponding
single-scale propagator $\dot{G}_\Lambda ( \bd{q} )$ defined in Eq.~(\ref{eq:Gdot}),
and the cross inside a loop means that all legs of the loop should subsequently be replaced by a single-scale propagator as defined in Eq.~(\ref{eq:productrule}).
}
\label{fig:flowh}
\end{figure}

\begin{figure}
 \begin{center}
  \centering
 \includegraphics[width=\columnwidth]{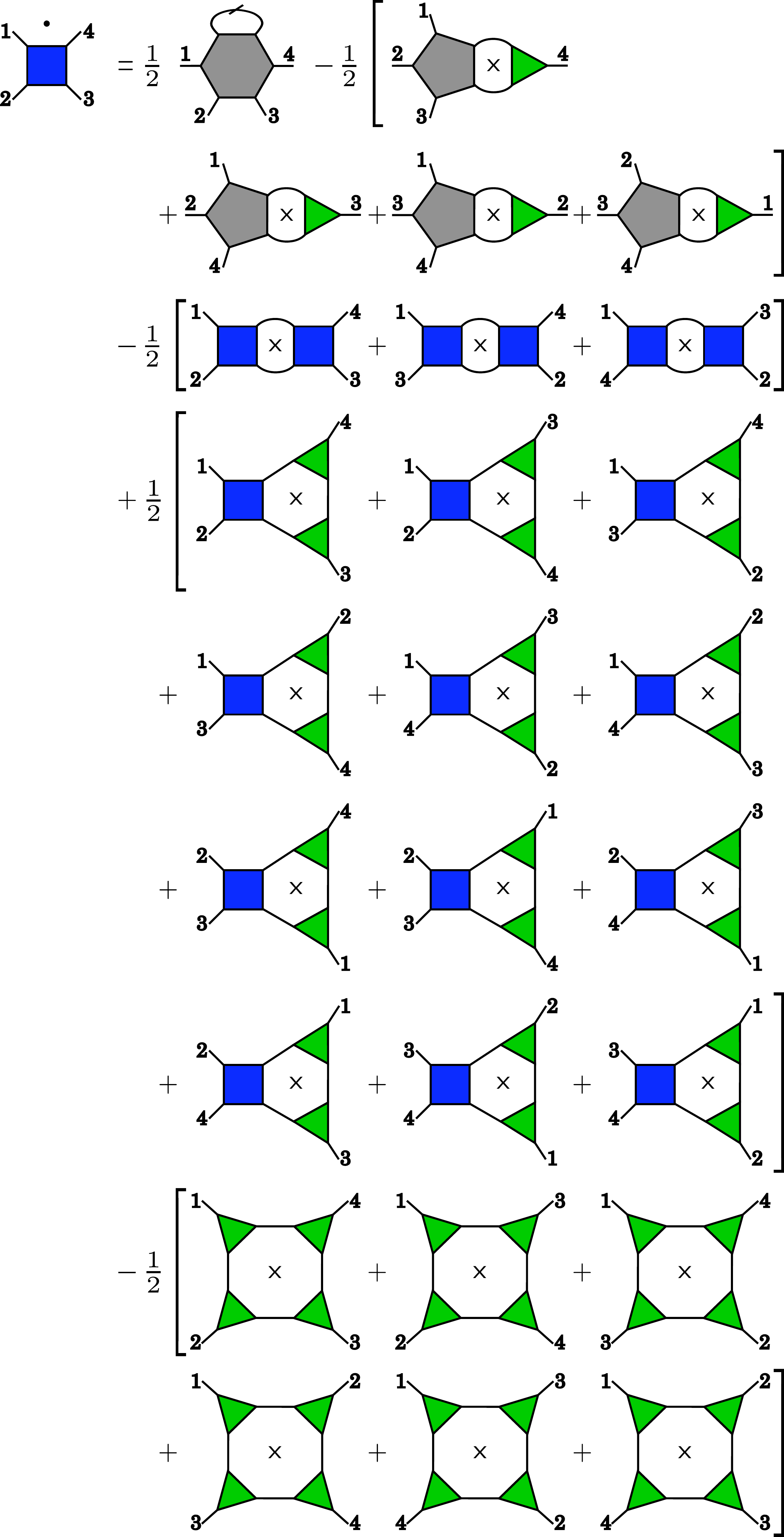}
   \end{center}
  \caption{%
Graphical representation of the exact FRG flow equation for the four-point vertex for constant order parameter $m$. The notation is the same as in Fig.~\ref{fig:flowh}. The five-point vertex and the the six-point vertex are represented by grey pentagons and hexagons, respectively.}
\label{fig:flowh4}
\end{figure}

By substituting the vertex expansion (\ref{eq:Gammashiftexp}) into the
Wetterich equation (\ref{eq:Wetterich}) and comparing symmetrized powers of the fields on both sides we obtain in infinite hierarchy of flow equations for the vertices. The zeroth order vertex can be identified with the Gibbs potential in units of temperature. It satisfies
\begin{equation}
 \partial_\Lambda {\Gamma}_{\Lambda}^{(0)} = 
 \frac{ \cal{V}}{2} \int_{\bd{k}} {G}_{\Lambda} ( \bd{k} ) \partial_{\Lambda}
  R_{\Lambda} (\bd{k} ),
  \label{eq:flowGamma0}
 \end{equation}
where
 \begin{align}
 G_{\Lambda} ( \bd{k} ) = \frac{1}{ {\Gamma}^{(2)}_{\Lambda} ( - \bd{k}, \bd{k} ) +  R_{\Lambda} ( \bd{k} ) 
  }
  \label{eq:Gdef}
 \end{align}
is the regularized two-point function. The flow of the one-point vertex
$\Gamma^{(1)}_{\Lambda} ( \bd{k}=0) = h_{\Lambda}$ is driven by the three-point vertex,
  \begin{align}
 \partial_{\Lambda} h_{\Lambda}
  & = 
   \frac{1}{2} \int_{\bd{k}} \dot{G}_{\Lambda} ( \bd{k} ) {\Gamma}^{(3)}_{\Lambda}
  ( - \bd{k} , \bd{k} , 0 ),
  \label{eq:hflow}
  \end{align}
where the so-called single-scale propagator is defined as
\begin{equation}
 \dot{G}_{\Lambda} ( \bd{k} ) = - G^2_{\Lambda} ( \bd{k} ) \partial_\Lambda
 R_{\Lambda} ( \bd{k} ).
 \label{eq:Gdot}
 \end{equation}
The two-point vertex satisfies the exact flow equation
\begin{widetext}
 \begin{align}
 &  \partial_\Lambda {\Gamma}^{(2)}_{\Lambda} ( - \bd{k} , \bd{k} )  =
 \frac{1}{2} \int_{\bd{q}} \dot{G}_{\Lambda} ( \bd{q} ){\Gamma}^{(4)}_{\Lambda}
  ( - \bd{q} , \bd{q} ,  - \bd{k} , \bd{k} )
 - \frac{1}{2} \int_{\bd{q}} [  {G}_{\Lambda} ( \bd{q} ) G_{\Lambda} ( \bd{q} + \bd{k} ) ]^{\bullet}
 \;
 {\Gamma}^{(3)}_{\Lambda} ( - \bd{q} , - \bd{k} , \bd{q} + \bd{k} )
  {\Gamma}^{(3) }_{\Lambda} (  \bd{q} ,  \bd{k} , - \bd{q} - \bd{k} ),
  \label{eq:flowGamma2m}
  \end{align}
while the RG flow of the three-point vertex is
 \begin{align}
 & \partial_{\Lambda}{\Gamma}^{(3)}_{\Lambda} ( \bd{k}_1 , \bd{k}_2, \bd{k}_3 )
  =  
 \frac{1}{2} \int_{\bd{q}} \dot{G}_{\Lambda} ( \bd{q} ) {\Gamma}^{(5)}_{\Lambda}
  ( - \bd{q} , \bd{q} ,  \bd{k}_1 , \bd{k}_2 , \bd{k}_3 )
 \nonumber
 \\
 & - \frac{1}{2} \int_{\bd{q}} \left\{
  [ {G}_{\Lambda} ( \bd{q} ) G_{\Lambda} ( \bd{q} + \bd{k}_3 ) ]^{\bullet} {\Gamma}^{(4)}_{\Lambda} ( \bd{k}_1, \bd{k}_2 , \bd{k}_3 + \bd{q} , - \bd{q} ){\Gamma}^{(3)}_{\Lambda} ( \bd{q} , -  \bd{q} - \bd{k}_3 , \bd{k}_3 )
   + {\small{ \begin{pmatrix}
  \bd{k}_1 \rightarrow \bd{k}_2 \\ \bd{k}_2 \rightarrow \bd{k}_3 \\
  \bd{k}_3 \rightarrow \bd{k}_1 \end{pmatrix} 
  + \begin{pmatrix}
  \bd{k}_1 \rightarrow \bd{k}_3 \\ \bd{k}_2 \rightarrow \bd{k}_1 \\
  \bd{k}_3 \rightarrow \bd{k}_2 \end{pmatrix} }}
 \right\}
\nonumber
\\
& +  \int_{\bd{q}} 
  [ {G}_{\Lambda} ( \bd{q} ) G_{\Lambda} ( \bd{q} + \bd{k}_1 )  G_{\Lambda} ( \bd{q} - \bd{k}_2 )]^{\bullet}
{\Gamma}^{(3)}_{\Lambda} ( \bd{k}_1, \bd{q} , - \bd{q} - \bd{k}_1 )  
 {\Gamma}^{(3)}_{\Lambda} ( \bd{k}_2, - \bd{q} ,  \bd{q} - \bd{k}_2 )  
{\Gamma}^{(3)}_{\Lambda} ( \bd{k}_3, \bd{q} + \bd{k}_1,  - \bd{q} + \bd{k}_2 )  .
\label{eq:flowGamma3m}
 \end{align}
Here, we have introduced the abbreviations
 \begin{subequations}
 \label{eq:productrule}
\begin{align}
 [  {G}_{\Lambda} ( \bd{k}_1 ) G_{\Lambda} ( \bd{k}_2 ) ]^{\bullet}
 & = \dot{G}_{\Lambda} ( \bd{k}_1 ) G_{\Lambda} ( \bd{k}_2 ) +
{G}_{\Lambda} ( \bd{k}_1 ) \dot{G}_{\Lambda} (  \bd{k}_2 ),
 \\
 [  {G}_{\Lambda} ( \bd{k}_1 ) G_{\Lambda} ( \bd{k}_2 ) G_{\Lambda} ( \bd{k}_3 )]^{\bullet}
 & = \dot{G}_{\Lambda} ( \bd{k}_1 ) G_{\Lambda} ( \bd{k}_2 )   
 G_{\Lambda} ( \bd{k}_3 ) +
{G}_{\Lambda} ( \bd{k}_1 ) \dot{G}_{\Lambda} (  \bd{k}_2 )  G_{\Lambda} ( \bd{k}_3 )  + {G}_{\Lambda} ( \bd{k}_1 ) {G}_{\Lambda} (  \bd{k}_2 )  \dot{G}_{\Lambda} ( \bd{k}_3 ).
\end{align}
\end{subequations}
 \end{widetext}
Graphical representations of the above flow equations are shown
in Fig.~\ref{fig:flowh}.
In the last line of Eq.~(\ref{eq:flowGamma3m}) we have used the fact that
the two diagrams involving three powers of the three-point vertex in the
last line of Fig.~\ref{fig:flowh} give identical contributions, which is easily seen be renaming $\bd{k}_1 \leftrightarrow \bd{k}_2$ and using $G_{\Lambda} ( \bd{q} ) = G_{\Lambda} ( - \bd{q} )$.
The exact flow equation for the four-point vertex (which is represented in Fig.~\ref{fig:flowh} by a blue square with four external legs)
is even more lengthy
because in a magnetic field it contains various combinations of vertices involving an odd number of external legs. A graphical representation of the related flow equation for constant magnetic field  
using a slightly different notation  can be found in Fig.~1 (d) of  Ref.~[\onlinecite{Schuetz06}].
To make this work self-contained, we show in Fig.~\ref{fig:flowh4} a graphical representation  the corresponding flow equation for constant $m$ using the same notation as in
Fig.~\ref{fig:flowh}. The combinatorial factors follow in a straightforward way from the
symmetrization operator in the general vertex expansion \cite{Kopietz10}.

To first order in $\varepsilon = 4 - D$ it is sufficient to approximate the two-point vertex by
 \begin{align}
 \Gamma^{(2)}_{\Lambda} ( - \bd{k} , \bd{k} ) & \approx
 c_0 k^2 +  r_{\Lambda},
 \end{align}
which  amounts to 
neglecting the momentum dependence of the self-energy. Note that
for small $\varepsilon $ the critical exponent $\eta$ is of order  $ \varepsilon^2 $ at the Ising fixed point, so that to first order in $\varepsilon$ it is consistent 
to approximate $\eta \approx 0$.
Similarly, we can also neglect the momentum dependence of the
three-point and four-point vertices,
 \begin{align}
 {\Gamma}^{(3)}_{\Lambda} ( \bd{k}_1 , \bd{k}_2 , \bd{k}_3  ) & \approx  {\Gamma}^{(3)}_{\Lambda} ( 0,0,0) = g_{\Lambda},
   \\
 {\Gamma}^{(4)}_{\Lambda} ( \bd{k}_1 , \bd{k}_2 , \bd{k}_3 , \bd{k}_4 ) & \approx {\Gamma}^{(4)}_{\Lambda} ( 0,0,0,0 ) = u_{\Lambda}.
  \end{align}
Moreover, we neglect all higher-order vertices, which is again justified  for small $\varepsilon = 4-D$. 
Our flow equation (\ref{eq:hflow}) for the scale-dependent magnetic field then
reduces to
\begin{align}
 \partial_{\Lambda} h_{\Lambda}  & =   \frac{g_{\Lambda}}{2} 
 \int_{\bd{q}}   \dot{G}_{\Lambda} ( \bd{q} ) ,
 \label{eq:hflow}
 \end{align}
while the flow equations for the other  three coupling constants are within our truncation given by
\begin{subequations}
  \label{eq:flowmcon}
 \begin{align}
  \partial_{\Lambda} r_{\Lambda} = {} &
    \frac{u_{\Lambda}}{2}
   \int_{\bd{q}}   \dot{G}_{\Lambda} ( \bd{q} )
  - \frac{g_{\Lambda}^2}{2} \int_{\bd{q}}  [ {G}^2_{\Lambda} ( \bd{q} ) 
    ]^{\bullet},
 \\
 \partial_{\Lambda} g_{\Lambda} = {} &
 - \frac{3}{2} g_{\Lambda} u_{\Lambda}  \int_{\bd{q}}  [ {G}^2_{\Lambda} ( \bd{q} ) ]^{\bullet}
 + g_{\Lambda}^3  \int_{\bd{q}}  [ {G}^3_{\Lambda} ( \bd{q} ) ]^{\bullet} , 
 \\
 \partial_{\Lambda} u_{\Lambda} = {} & - \frac{3}{2}  u_{\Lambda}^2  
 \int_{\bd{q}}  [ {G}^2_{\Lambda} ( \bd{q} ) ]^{\bullet}
 + 6 u_{\Lambda} g_{\Lambda}^2  \int_{\bd{q}}  [ {G}^3_{\Lambda} ( \bd{q} ) ]^{\bullet}
 \nonumber
 \\
 & -  3 g_{\Lambda}^4  \int_{\bd{q}}  [ {G}^4_{\Lambda} ( \bd{q} ) ]^{\bullet}.
 \label{eq:uLambdaflowm}
 \end{align}
 \end{subequations}
These flow equations should be integrated with the initial conditions
 \begin{subequations}
 \label{eq:initialvalues}
 \begin{align}
 \Gamma_{\Lambda_0}^{(0)} & = {\cal{V}} \left( \frac{{r}_0}{2} m^2 + \frac{ u_0}{ 4!} m^4 \right),
 \\
 \Gamma_{\Lambda_0}^{(1)} ( \bd{k} =0) & = 	h_0 = r_0 m + \frac{u_0}{3!} m^3,
 \\
 \Gamma^{(2)}_{\Lambda_0} ( - \bd{k} , \bd{k} ) & = c_0 k^2 + r_{\Lambda_0} = 
  c_0 k^2 + r_0 + \frac{u_0}{2} m^2,
 \\
 g_{\Lambda_0} & = u_0 m ,
 \end{align} 
 \end{subequations}
which define the initial values of the vertices in terms of the fixed magnetization $m$.
Note that at the initial scale the value of the mass parameter for finite $m$ is
 \begin{equation}
 r_{\Lambda_0} = r_0 + \frac{u_0}{2} m^2.
 \end{equation}

\subsection{Scaling at the Ising fixed point for constant $m$}
\label{sec:scalingh}

\begin{figure}
 \begin{center}
  \centering
 \includegraphics[width=1.14\columnwidth]{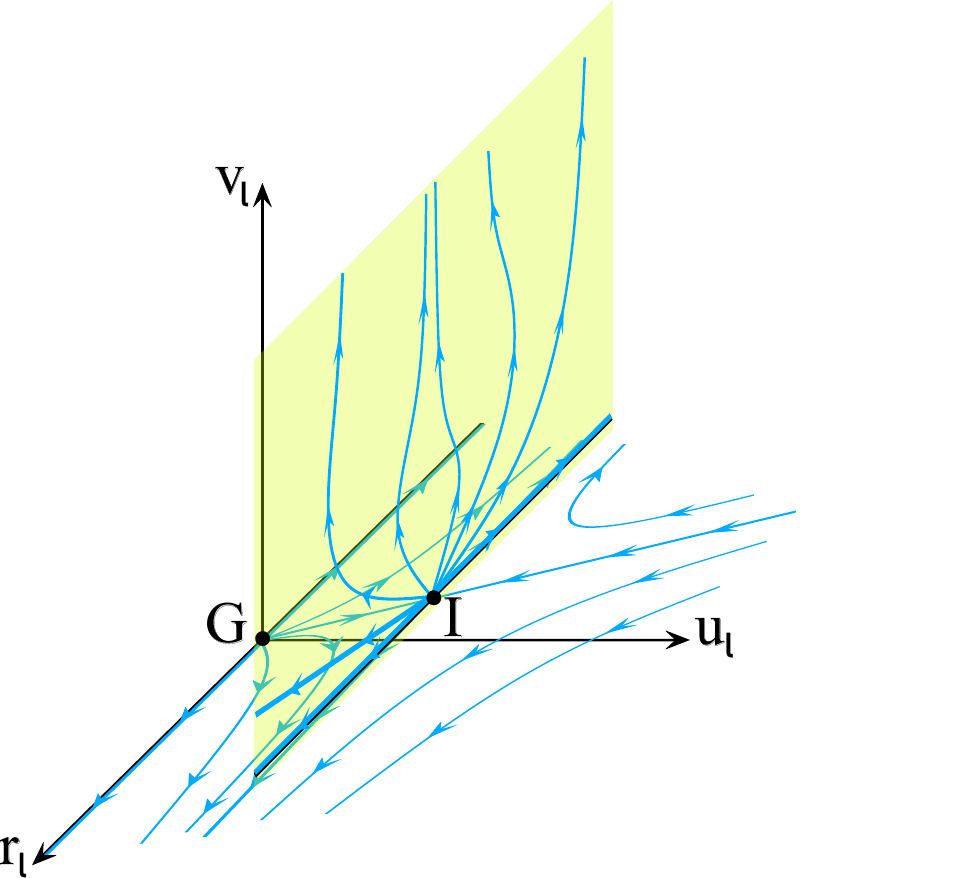}
   \end{center}
  \caption{%
Qualitative behavior of the RG flow of $\phi^4$ theory for constant  magnetization
in the truncated coupling space  spanned by $r_l$, $u_l$ and $v_l = g_l^2$ [see
Eqs.~(\ref{eq:rlflowm}--\ref{eq:ulflowm})] for small $\varepsilon = 4-D > 0$.
There are two fixed points:
the unstable Gaussian (G) fixed point and the non-trivial Ising (I) fixed point. 
The thick blue lines represent the eigendirections of the flow.
The green surface highlights the flow in the $ r_l $-$ v_l $ plane containing the Ising fixed point I. 
}
\label{fig:RGflowm}
\end{figure}

For simplicity we now use a sharp momentum cutoff. Within our truncation the regularized two-point function is then given by
 \begin{equation}
 G_{\Lambda} ( \bd{q} ) = \frac{\Theta ( | \bd{q} | - \Lambda )}{
 c_0 \bd{q}^2 + \Theta ( | \bd{q} | - \Lambda ) r_{\Lambda} }.
 \end{equation}
For the evaluation of the 
corresponding single-scale propagator defined via (\ref{eq:Gdot}) 
and the momentum integrals in Eq.~(\ref{eq:flowmcon})
we should carefully treat the relevant 
combinations of $\delta$- and $\Theta$-functions 
using \cite{Morris94,Kopietz10}
 \begin{equation}
 \delta ( x ) f ( \Theta (x ) ) = \delta ( x ) \int_0^1 d t f ( t ),
 \end{equation}
where $f ( \Theta )$ is an arbitrary function of $\Theta ( x )$.
The resulting single-scale propagator for sharp momentum cutoff is simply
 \begin{equation}
  \dot{G}_{\Lambda} ( \bd{q} ) = - \frac{\delta ( | \bd{q} | - \Lambda )}{c_0 \Lambda^2 + r_{\Lambda} },
  \end{equation}
while our  flow equation (\ref{eq:flowGamma0})
for the Gibbs potential reduces to
 \begin{align}
 \partial_{\Lambda} \Gamma^{(0)}_{\Lambda} & = - \frac{\cal{V}}{2} \int_{\bd{q}} \delta ( | \bd{q} | - \Lambda ) \ln \left( \frac{ c_0 \Lambda^2 + r_{\Lambda} }{ c_0 \Lambda^2 } \right)
 \nonumber
 \\
 & = - {\cal{V}}   \frac{  K_D }{2} \Lambda^{D-1} \ln \left( \frac{ c_0 \Lambda^2 + r_{\Lambda} }{ c_0 \Lambda^2 } \right).
 \label{eq:flowGamma0sharp}
 \end{align}
Here the numerical constant $K_D$ is given by the surface area of the
$D$-dimensional unit sphere divided by $(2 \pi )^D$.
The flow equation (\ref{eq:hflow}) for the renormalized magnetic field
reduces to
 \begin{equation}
 \partial_{\Lambda} h_{\Lambda}  =  - \frac{K_D}{2} \frac{ g_{\Lambda} \Lambda^{D-1}}{ c_0 \Lambda^2 + r_{\Lambda} },
 \label{eq:flowhsharp}
 \end{equation}
while the flow  equations (\ref{eq:flowmcon}) for the remaining couplings can be written as
 \begin{subequations}
 \label{eq:flowrgusharp}
 \begin{align}
 \partial_{\Lambda} r_{\Lambda} = {} & - \frac{ K_D}{2} \frac{ u_{\Lambda} \Lambda^{D-1} }{ c_0 \Lambda^2 + r_{\Lambda} } 
  + \frac{K_D}{2} \frac{ g_{\Lambda}^2 \Lambda^{D-1}}{ ( c_0 \Lambda^2 + r_{\Lambda} )^2 }, 
 \\
 \partial_{\Lambda} g_{\Lambda} = {} & 
  \frac{ 3 K_D}{2}  \frac{ g_{\Lambda} u_{\Lambda} \Lambda^{D-1} }{ ( c_0 \Lambda^2 + r_{\Lambda} )^2 } 
 - K_D \frac{  g^3_{\Lambda} \Lambda^{D-1} }{ ( c_0 \Lambda^2 + r_{\Lambda} )^3 } ,
\\   
\partial_{\Lambda} u_{\Lambda} = {} &
  \frac{ 3 K_D}{2}  \frac{ u^2_{\Lambda} \Lambda^{D-1} }{ ( c_0 \Lambda^2 + r_{\Lambda} )^2 } 
 - 6 K_D   \frac{ u_{\Lambda} g^2_{\Lambda} \Lambda^{D-1} }{ ( c_0 \Lambda^2 + r_{\Lambda} )^3 }
 \nonumber
 \\
  & + 3 K_D \frac{  g^4_{\Lambda} \Lambda^{D-1} }{ ( c_0 \Lambda^2 + r_{\Lambda} )^4 } .
  \end{align}
 \end{subequations}
To calculate the fixed points of the RG flow and  associated scaling properties we now introduce the following dimensionless rescaled couplings:
\begin{subequations}
\label{eq:couplingshscale}
  \begin{align}
   f_l & = \frac{ \Gamma^{(0)}_{\Lambda}}{ {\cal{V}} K_D \Lambda^D}
   \\
   h_l & = \sqrt{ \frac{ 1}{K_D c_0 \Lambda^{D+2}}}  h_{\Lambda},
   \\
  r_l & = \frac{ r_{\Lambda}}{ c_0 \Lambda^2} ,
  \\
  g_l & = \sqrt{ \frac{ K_D \Lambda^{ D-6}}{c_0^3} } g_{\Lambda},
  \\
  u_l &  = \frac{ K_D  \Lambda^{D-4}  }{c_0^2}  u_{\Lambda},
 \end{align}
 \end{subequations}
which are functions of the  logarithmic flow parameter
$l = \ln ( \Lambda_0 / \Lambda )$. Then we obtain the following rescaled version 
of the flow equations (\ref{eq:flowGamma0sharp}--\ref{eq:flowrgusharp}):
 \begin{subequations}
 \label{eq:allflowm}
 \begin{align}
 \partial_l f_l & = D f_l + \frac{ 1}{2} \ln ( 1 + r_l ),
  \label{eq:flflow}
  \\
 \partial_l h_l & = \frac{ D+2 }{2} h_l + \frac{ g_l}{2 ( 1 + r_l ) },
 \label{eq:hlflow}
 \\
 \partial_l r_l & = 2 r_l + \frac{ u_l}{2 ( 1 + r_l ) }
 - \frac{ g_l^2}{ 2 (1 + r_l )^2} ,
 \label{eq:rlflowm}
 \\
 \partial_l g_l & = \frac{ 6-D}{2} g_l - \frac{ 3 g_l u_l }{2 ( 1 + r_l )^2 }
  + \frac{ g_l^3}{ ( 1 + r_l )^3 },  
  \label{eq:glflowm}
 \\
 \partial_l u_l & = ( 4 - D ) u_l - \frac{ 3 u_l^2}{ 2 ( 1 + r_l )^2}
 +  \frac{ 6 u_l g_l^2}{ ( 1 + r_l )^3 } -  \frac{ 3 g_l^4}{ ( 1 + r_l )^4 }.
 \label{eq:ulflowm}
  \end{align} 
 \end{subequations}
A sketch of the RG flow in the
space of couplings spanned by $r_l$, $u_l$ and $v_l = g_l^2$
slightly below four dimensions 
is shown Fig.~\ref{fig:RGflowm}.
For $ h_l = g_l=0$ Eqs.~(\ref{eq:rlflowm}) and
(\ref{eq:ulflowm})  reduce to the usual
one-loop flow equations for $\phi^4$ theory in the absence of a magnetic field \cite{Ma76,Kopietz10},
 \begin{subequations}
 \label{eq:ruflow}
 \begin{align}
 \partial_l r_l & = 2 r_l + \frac{ u_l}{2 ( 1 + r_l ) },
 \\
\partial_l u_l & = ( 4 - D ) u_l - \frac{ 3 u_l^2}{ 2 ( 1 + r_l )^2}.
 \end{align}
 \end{subequations}
With $\varepsilon = 4-D$ the fixed point conditions are
 \begin{subequations}
 \begin{align}
 2 r_{\ast} & = - \frac{ u_{\ast}}{2 ( 1 + r_{\ast} )}, 
 \label{eq:rast}
 \\
 \varepsilon u_{\ast} & = \frac{ 3 u_\ast^2}{ 2 ( 1 + r_\ast )^2}.
 \label{eq:uast}
 \end{align}
 \end{subequations}
For $u_{\ast} \neq 0$ Eq.~(\ref{eq:uast}) can be written as
  \begin{equation}
 \frac{ u_{\ast} }{ 1 + r_{\ast} } = \frac{2}{3} \varepsilon ( 1 + r_{\ast} ).
  \end{equation}  
Substituting this into Eq.~(\ref{eq:rast}) we obtain
 \begin{equation}
  r_{\ast} = - \frac{ \varepsilon}{6} ( 1 + r_{\ast} ),
  \end{equation}   
and hence
 \begin{equation}
 r_{\ast} = - \frac{\varepsilon}{ 6 + \varepsilon} =  - \frac{\varepsilon}{6} + {\cal{O}} ( \varepsilon^2 ) .
 \end{equation} 
Substituting this back into Eq.~(\ref{eq:rast}) and solving for $u_{\ast}$ we obtain  the value of the dimensionless two-body interaction 
at the Ising fixed point,
 \begin{align}
 u_{\ast} & = - 4 r_{\ast} ( 1 + r_{\ast} ) = \frac{ \frac{ 2 \varepsilon}{3} }{ 
 \left( 1 + \frac{ \varepsilon}{6} \right)^2 } =
 \frac{ 2 \varepsilon}{3} + {\cal{O}} ( \varepsilon^2 ).
 \end{align}
Because in our
truncation we have neglected the anomalous dimension $\eta = {\cal{O}} ( \varepsilon^2 )$
it is consistent to expand the couplings at the fixed point to first order in~$\varepsilon$.

To  calculate the scaling in the vicinity  of the 
Ising
fixed point for constant magnetization, 
we linearize the RG flow
around the fixed point. Setting $r_l = r_{\ast} + \delta r_l$, $u_l = u_{\ast} + \delta u_l$, and
$v_l = g_l^2$ we obtain to first order in $\varepsilon$,
 \begin{align}
 \partial_l \begin{pmatrix} \delta r_l \\ \delta u_l \\ v_l \end{pmatrix}  =
 \begin{pmatrix} 2 - \frac{\varepsilon}{3} & \frac{1}{2} +
  \frac{\varepsilon}{12} & -\frac{1}{2} - \frac{\varepsilon}{6} \\
  0 & - \varepsilon & 4 \varepsilon \\
  0 & 0 & 2 - \varepsilon \end{pmatrix}   \begin{pmatrix} \delta r_l \\ \delta u_l \\ v_l \end{pmatrix} .
  \label{eq:flowlinm}
 \end{align}
Since the coefficient matrix is upper triangular, its eigenvalues can be identified  from the diagonal,
 \begin{subequations}
 \begin{align}
 y_t & = 2 - \frac{ \varepsilon}{3} = \frac{ D+2}{3} ,
 \\
 y_s & = - \varepsilon,
 \\
 y_v & = 2 - \varepsilon = D -2 .
 \label{eq:yvm}
\end{align}
 \end{subequations}
To determine the corresponding scaling variables \cite{Ma76,Kopietz10},
we need the left-eigenvectors, which to first order in
$\varepsilon$ are 
 \begin{subequations}
\begin{align}
 \bd{\psi}_t^T &  = 
\left(  - \frac{  4 \varepsilon }{ 3 }, 
   - \frac{ \varepsilon }{ 3  }, \; 1   \right) ,
\\
 \bd{\psi}_s^T & = \left( 0 , 1 , - 2 \varepsilon\right),
 \\
 \bd{\psi}_v^T & = ( 0,0,1 ),
\end{align}
 \end{subequations}
where we have chosen a convenient normalization.
For small $\varepsilon$ the corresponding scaling variables are given by
 \begin{subequations}
 \label{eq:tsvdef}
  \begin{align}
 t_l & =  \bd{\psi}_t^T \begin{pmatrix} \delta r_l \\ \delta u_l \\ v_l \end{pmatrix} =  - \frac{4 \varepsilon}{3} \delta r_l  - \frac{ \varepsilon}{3}  \delta u_l 
 +  v_l ,
 \label{eq:tscale}
 \\
 s_l & = \bd{\psi}_s^T \begin{pmatrix} \delta r_l \\ \delta u_l \\ v_l \end{pmatrix} = \delta u_l  - 2 \varepsilon  v_l,
 \label{eq:sscale}
 \\
 v_l & = \bd{\psi}_v^T \begin{pmatrix} \delta r_l \\ \delta u_l \\ v_l \end{pmatrix} .
 \end{align}
 \end{subequations}
By construction, in terms of the  scaling variables the linearized flow around the fixed points decouples,
 \begin{subequations}
 \begin{align}
 \partial_l t_l &  = y_t t_l,
 \\
 \partial_l s_l & = y_s s_l,
 \\
 \partial_l v_l & = y_v v_l,
 \end{align}
 \end{subequations}
so that
 \begin{subequations}
 \begin{align}
 t_l & = t_0 e^{ y_t l }  = t_0 e^{ \frac{( D+2)}{3} l } =  t_0 e^{ (2 - \varepsilon/3) l },
 \\
 s_l & = s_0 e^{y_s l} = s_0 e^{ -( 4 -D ) l } = s_0 e^{ - \varepsilon l },
 \\
 v_l & = v_0 e^{ y_v l } = v_0 e^{ ( D - 2 ) l } = v_0 e^{ (2 - \varepsilon ) l }.
 \end{align}
 \end{subequations} 
Note that according to Eq.~(\ref{eq:tscale}) 
the thermal eigenvector $\bd{\psi}^T_t$ 
is  for small $\varepsilon$ almost parallel to the magnetic eigenvector $\bd{\psi}^T_v$.
%
Hence, for small $\varepsilon$ the
thermal scaling variable $t_l$ that can be controlled by changing the temperature via  the initial  value of $\delta r_0$
is dominated by the magnetic component $v_l$. As a consequence, if 
the magnetization and the temperature are fine-tuned
to the critical point, then the fine-tuning is more sensitive to the magnetization than to the temperature.

To derive the magnetic equation of state at the critical temperature, 
we assume that the temperature is adjusted  such  that $t_0 =0$.
Keeping in mind that for small magnetization $m$ the scaling variable $v_l = g_l^2$ is
proportional to $m^2$, the singular part of the Gibbs potential $\Gamma_{\rm sing} (m) $ for constant magnetization  satisfies the scaling law
 \begin{equation}
\Gamma_{\rm sing} ( m )  =   b^{-D} \Phi ( b^{ y_v} m^2 ), 
 \end{equation}
where the scaling factor $b$ is arbitrary and $\Phi ( x )$ is the relevant  scaling function.
Setting $ b = m^{ -2 / y_v }$ we obtain
 \begin{equation}
 \Gamma_{\rm sing} ( m ) =   m^{ 2 D / y_v } \Phi (1 )  \propto m^{ 2 D / y_v }.
 \end{equation}
The magnetic equation of state at the critical temperature is therefore
 \begin{equation}
  h \propto \frac{ \partial \Gamma_{\rm sing} (m) }{\partial m } \propto m^{ \frac{2 D}{y_v} -1 }
  = m^{ \delta },
  \end{equation}
with critical exponent  
 \begin{equation}
 \delta = \frac{ 2 D}{y_v} -1  = \frac{ D+2}{D-2} = 3 + \varepsilon +
 {\cal{O}} ( \varepsilon^2 ).
 \label{eq:deltares}
 \end{equation}
Note that for $D < 4$ where hyperscaling is satisfied the exact exponent $\delta$ 
is related to the correlation function exponent $\eta$ via the scaling relation  \cite{Ma76,Fisher83}
\begin{equation}
 \delta = \frac{ D+2 - \eta}{D-2 + \eta}.
 \end{equation}
Keeping in mind that $\eta = {\cal{O}} ( \varepsilon^2 )$ 
we see that our result (\ref{eq:deltares}) is correct to order $\varepsilon$.
The scaling function $\Phi  (x )$ and the scaling form of the critical equation of state  can in principle be obtained by integrating
the flow equations (\ref{eq:flflow}) and (\ref{eq:hlflow}),
 \begin{align}
  f_l & = f_{l=0} e^{Dl} + \int_0^l d t e^{D ( l -t )} \frac{1}{2} \ln ( 1 + r_t ) ,
  \label{eq:flint}
  \\
  h_l & = h_{l=0} e^{ \frac{D+2}{2} l } + 
  \int_0^l d t e^{\frac{D+2}{2} ( l -t )} \frac{g_t}{2 ( 1 + r_t )} .
  \label{eq:hlint}
 \end{align}
For $l \rightarrow \infty$ we obtain from Eq.~(\ref{eq:flint})
for the Gibbs potential  for constant magnetization
 \begin{align}
 \Gamma( m ) & = {\cal{V}} K_D \lim_{l \rightarrow \infty} \Lambda^D f_l
 \nonumber
 \\
 &  = {\cal{V}} K_D \Lambda_0^D \left[ f_{l=0} + \int_0^{\infty} d t e^{ - D t }
  \frac{1}{2} \ln ( 1 + r_t )  \right],
 \end{align}
and for the magnetic equation of state
 \begin{align}
& h ( m )  =   \lim_{ l \rightarrow \infty} \sqrt{ K_D c_0 \Lambda^{D+2} }
  h_l \nonumber
  \\
  & = \sqrt{ K_D c_0 \Lambda_0^{D+2} }
  \left[ h_{l=0} + \int_0^\infty d t e^{ - \frac{D+2}{2} t} \frac{g_t}{2 ( 1 + r_t )} \right] .
 \end{align}
At the critical point the initial values of the relevant couplings have be to fine-tuned such that the coefficients of order $m$ and  $m^3$ in the expansion of $h ( m )$ vanish;  the leading term is then non-analytic,  $h \propto m^{\delta}$, with $\delta = 3 + \varepsilon +
{\cal{O}} ( \varepsilon^2 )$.

\section{FRG for constant magnetic field}
\label{sec:RGh}

\subsection{Flow equations for constant $h$}
When the magnetic field $ h = h_0$ is fixed during the RG procedure, the magnetization
$m_{\Lambda}$ necessarily depends on the RG scale $\Lambda$. 
The fundamental equations which determine the $\Lambda$-dependence of $m_{\Lambda}$ are
\begin{equation}
 \left. \frac{ \delta {\cal{G}}_{\Lambda} [ J ] }{\delta J ( \bd{r})  }
  \right|_{ J  ( \bd{r} ) =h} = m_{\Lambda},
 \label{eq:Ghm2}
 \end{equation}
and
\begin{equation}
 \left. \frac{ \delta {\Gamma}_{\Lambda} [ \bar{\phi} ] }{\delta 
 \bar{\phi}  ( \bd{r})  }
  \right|_{ \bar{\phi} ( \bd{r} ) =m_{\Lambda}}
  =  h,
  \label{eq:Gammamh2}
\end{equation}
which should be compared with the corresponding equations~(\ref{eq:Ghm1}) and
(\ref{eq:Gammamh1}) for constant $m$.
Shifting $\bar{\phi}  ( \bd{r} ) = m_{\Lambda} + \varphi ( \bd{r} ) $ 
and defining
 \begin{equation}
  \tilde{\Gamma}_\Lambda [ \varphi ] = \Gamma_{\Lambda} [ m_{\Lambda} + \varphi  ],
  \label{eq:Gammashift}
  \end{equation}
the condition (\ref{eq:Gammamh2}) can also be written as
 \begin{equation}   
   \left.  
 \frac{ \delta \tilde{\Gamma}_{\Lambda} [ \varphi ]}{\delta {\varphi} (\bd{r})} \right|_{  {\varphi} ( \bd{r} )  = 0}  = h.
  \label{eq:extremal2}
 \end{equation}
Using the fact that
the functional  $\Gamma_{\Lambda} [ \bar{\phi}  ]$ satisfies the Wetterich 
equation (\ref{eq:Wetterich}) we conclude that
the shifted functional $\tilde{\Gamma}_{\Lambda} [ \varphi ]$ satisfies
\begin{align}
 \partial_{\Lambda} \tilde{\Gamma}_{\Lambda} [ {\varphi} ] = {} & 
 \frac{1}{2} {\rm Tr} \left[ ( \mathbf{\tilde{\Gamma}}^{\prime \prime}_{\Lambda} [  {\varphi} ] + \mathbf{R}_{\Lambda} )^{-1} \partial_\Lambda \mathbf{R}_{\Lambda} \right]
 \nonumber
  \\
 &   
   +  \frac{ \delta \tilde{\Gamma}_{\Lambda} [ \varphi] }{\delta \varphi_{ \bd{k} =0 } }        
    \partial_{\Lambda} m_{\Lambda},
    \label{eq:Wetterich2}
 \end{align}
where the last term is generated by the scale-dependence of $m_{\Lambda}$ in the shifted argument of $\Gamma_{\Lambda} [ m_{\Lambda} + \varphi ]$.
The condition (\ref{eq:extremal2}) implies that in momentum space the vertex expansion 
of the functional $\tilde{\Gamma}_{\Lambda} [ {\varphi} ]$
has the form,
 \begin{align}
  \tilde{\Gamma}_{\Lambda} [ \varphi ]   = \tilde{\Gamma}_{\Lambda}^{(0)}
  + &  \sum_{ n =1}^{\infty} \frac{1}{n!} \int_{\bd{k}_1} \ldots \int_{\bd{k}_n }
  {\cal{V}} \delta_{ \bd{k}_1 + \ldots + \bd{k}_n , 0 }
  \nonumber
   \\
   & \times 
   \tilde{\Gamma}^{(n)}_{\Lambda} ( \bd{k}_1 , \ldots , \bd{k}_n )
  \varphi_{ \bd{k}_1 } \ldots \varphi_{\bd{k}_n } ,
  \label{eq:vertexexpansion}
\end{align}
with the  boundary condition that the one-point vertex $\tilde{\Gamma}^{(1)}_{\Lambda} ( 
\bd{k} =0 )$ is fixed to the value of the external magnetic field \cite{remarkfunctional},
 \begin{equation}
 \tilde{\Gamma}^{(1)}_{\Lambda} ( \bd{k} =0 ) = h.
 \label{eq:Gamma1cond}
 \end{equation}
By inserting the vertex expansion (\ref{eq:vertexexpansion})
into the modified Wetterich equation (\ref{eq:Wetterich2}) and comparing both sides,
we obtain again an infinite hierarchy of flow equations for the vertices. 
The scale-dependence of the magnetization generates an extra term in the flow 
equation of all vertices \cite{Kopietz10,Schuetz06}.
The field-independent part  is
 \begin{equation}
 \partial_\Lambda \tilde{\Gamma}^{(0)}_{\Lambda} = 
 {\cal{V}} h \partial_{\Lambda} m_{\Lambda} + 
 \frac{ \cal{V}}{2} \int_{\bd{k}} {G}_{\Lambda} ( \bd{k} ) \partial_{\Lambda}
  R_{\Lambda} (\bd{k} ),
  \label{eq:flowfreeenergy}
  \end{equation}
which contains the extra term ${\cal{V}} h \partial_{\Lambda} m_{\Lambda} $
in comparison with the corresponding flow equation (\ref{eq:flowGamma0}) for constant $m$.
Here the scale-dependent regulated two-point function is defined analogous to
Eq.~(\ref{eq:Gdef}),
 \begin{equation}
 G_{\Lambda} ( \bd{k} ) = \frac{1}{ \tilde{\Gamma}^{(2)}_{\Lambda} ( - \bd{k}, \bd{k} ) +  R_{\Lambda} ( \bd{k} ) 
  }.
 \end{equation}
From the condition (\ref{eq:Gamma1cond}) that
the one-point vertex is scale-independent we obtain the flow equation for the  magnetization,
 \begin{align}
 0 & = \tilde{\Gamma}^{(2)}_{\Lambda} ( 0,0) \partial_{\Lambda} m_{\Lambda}
  + \frac{1}{2} \int_{\bd{q}} \dot{G}_{\Lambda} ( \bd{q} ) \tilde{\Gamma}^{(3)}_{\Lambda}
  ( - \bd{q} , \bd{q} , 0 ),
  \label{eq:magflow}
  \end{align}
where the single-scale propagator
 $
 \dot{G}_{\Lambda} ( \bd{k} ) = - G^2_{\Lambda} ( \bd{k} ) \partial_\Lambda
 R_{\Lambda} ( \bd{k} )$
is defined as in Eq.~(\ref{eq:Gdot}).
The exact flow equation for the two-point vertex is now
 \begin{widetext}
 \begin{align}
   \partial_\Lambda \tilde{\Gamma}^{(2)}_{\Lambda} ( - \bd{k} , \bd{k} ) & =
 \tilde{\Gamma}^{(3)}_{\Lambda} ( - \bd{k} , \bd{k} ,0)   
 \partial_{\Lambda} m_{\Lambda} 
   + \frac{1}{2} \int_{\bd{q}} \dot{G}_{\Lambda} ( \bd{q} ) \tilde{\Gamma}^{(4)}_{\Lambda}
  ( - \bd{q} , \bd{q} ,  - \bd{k} , \bd{k} )
 \nonumber
 \\
 & - \frac{1}{2} \int_{\bd{q}} [  {G}_{\Lambda} ( \bd{q} ) G_{\Lambda} ( \bd{q} + \bd{k} ) ]^{\bullet}
 \;
 \tilde{\Gamma}^{(3)}_{\Lambda} ( - \bd{q} , - \bd{k} , \bd{q} + \bd{k} )
   \tilde{\Gamma}^{(3) }_{\Lambda} (  \bd{q} ,  \bd{k} , - \bd{q} - \bd{k} ),
  \end{align}
which differs from the corresponding flow equation (\ref{eq:flowGamma2m}) for constant $m$ by the  term $\tilde{\Gamma}^{(3)}_{\Lambda} ( - \bd{k} , \bd{k} ,0)   
 \partial_{\Lambda} m_{\Lambda} $ involving the scale-derivative of the magnetization.
Similarly, for the flow of the  three-point vertex at constant $h$ we obtain
 \begin{align}
 & \partial_{\Lambda} \tilde{\Gamma}^{(3)}_{\Lambda} ( \bd{k}_1 , \bd{k}_2, \bd{k}_3 )
  =  \tilde{\Gamma}^{(4)}_{\Lambda} ( \bd{k}_1 , \bd{k}_2, \bd{k}_3 ,0 ) 
  \partial_{\Lambda} m_{\Lambda} 
   + \frac{1}{2} \int_{\bd{q}} \dot{G}_{\Lambda} ( \bd{q} ) \tilde{\Gamma}^{(5)}_{\Lambda}
  ( - \bd{q} , \bd{q} ,  \bd{k}_1 , \bd{k}_2 , \bd{k}_3 )
 \nonumber
 \\
 & - \frac{1}{2} \int_{\bd{q}} \left\{
  [ {G}_{\Lambda} ( \bd{q} ) G_{\Lambda} ( \bd{q} + \bd{k}_3 ) ]^{\bullet} \tilde{\Gamma}^{(4)}_{\Lambda} ( \bd{k}_1, \bd{k}_2 , \bd{k}_3 + \bd{q} , - \bd{q} ) \tilde{\Gamma}^{(3)}_{\Lambda} ( \bd{q} , -  \bd{q} - \bd{k}_3 , \bd{k}_3 )
  + 
  {\small{ \begin{pmatrix}
  \bd{k}_1 \rightarrow \bd{k}_2 \\ \bd{k}_2 \rightarrow \bd{k}_3 \\
  \bd{k}_3 \rightarrow \bd{k}_1 \end{pmatrix} 
  + \begin{pmatrix}
  \bd{k}_1 \rightarrow \bd{k}_3 \\ \bd{k}_2 \rightarrow \bd{k}_1 \\
  \bd{k}_3 \rightarrow \bd{k}_2 \end{pmatrix} }}
 \right\}
\nonumber
\\
& +  \int_{\bd{q}} 
  [ {G}_{\Lambda} ( \bd{q} ) G_{\Lambda} ( \bd{q} + \bd{k}_1 )  G_{\Lambda} ( \bd{q} - \bd{k}_2 )]^{\bullet}
 \tilde{\Gamma}^{(3)}_{\Lambda} ( \bd{k}_1, \bd{q} , - \bd{q} - \bd{k}_1 )  
 \tilde{\Gamma}^{(3)}_{\Lambda} ( \bd{k}_2, - \bd{q} ,  \bd{q} - \bd{k}_2 )  
 \tilde{\Gamma}^{(3)}_{\Lambda} ( \bd{k}_3, \bd{q} + \bd{k}_1,  - \bd{q} + \bd{k}_2 )  ,
 \end{align}
 \end{widetext}
which replaces Eq.~(\ref{eq:flowGamma3m}).
As in Sec.~\ref{sec:RGm} we now approximate  
 \begin{subequations}
 \label{eq:rgudef2}
 \begin{align}
  \tilde{\Gamma}^{(2)}_{\Lambda} ( - \bd{k} , \bd{k} ) & \approx
     \tilde{\Gamma}^{(2)}_{\Lambda} ( 0 , 0 ) = r_{\Lambda},
    \\
  \tilde{\Gamma}^{(3)}_{\Lambda} ( \bd{k}_1 , \bd{k}_2 , \bd{k}_3  ) & \approx  \tilde{\Gamma}^{(3)}_{\Lambda} ( 0,0,0) = g_{\Lambda},
   \\
  \tilde{\Gamma}^{(4)}_{\Lambda} ( \bd{k}_1 , \bd{k}_2 , \bd{k}_3 , \bd{k}_4 ) & \approx  \tilde{\Gamma}^{(4)}_{\Lambda} ( 0,0,0,0 ) = u_{\Lambda}.
  \end{align}
  \end{subequations}
Then Eq.~(\ref{eq:magflow}) reduces to
 \begin{subequations}
  \label{eq:flowhcon}
 \begin{align}
 0 & = r_{\Lambda}  \partial_{\Lambda} m_{\Lambda} +  \frac{g_{\Lambda}}{2} 
 \int_{\bd{q}}   \dot{G}_{\Lambda} ( \bd{q} ) ,
 \end{align}
while the flow of the couplings defined in Eq.~(\ref{eq:rgudef2}) reduces to
  \begin{align}
  \partial_{\Lambda}  r_\Lambda = {} &
   g_{\Lambda} \partial_{\Lambda} m_{\Lambda}  + \frac{u_{\Lambda}}{2}
   \int_{\bd{q}}   \dot{G}_{\Lambda} ( \bd{q} )
    - \frac{g_{\Lambda}^2}{2} \int_{\bd{q}}  [ {G}^2_{\Lambda} ( \bd{q} ) 
    ]^{\bullet},
 \\
 \partial_{\Lambda} g_{\Lambda} = {} & u_{\Lambda} \partial_{\Lambda} m_{\Lambda}
 - \frac{3}{2} g_{\Lambda} u_{\Lambda}  \int_{\bd{q}}  [ {G}^2_{\Lambda} ( \bd{q} ) ]^{\bullet}
 \nonumber
 \\
 & + 
 g_{\Lambda}^3  \int_{\bd{q}}  [ {G}^3_{\Lambda} ( \bd{q} ) ]^{\bullet} , 
 \\
 \partial_{\Lambda} u_{\Lambda} = {} & - \frac{3}{2}  u_{\Lambda}^2  
 \int_{\bd{q}}  [ {G}^2_{\Lambda} ( \bd{q} ) ]^{\bullet}
 + 6 u_{\Lambda} g_{\Lambda}^2  \int_{\bd{q}}  [ {G}^3_{\Lambda} ( \bd{q} ) ]^{\bullet}
 \nonumber
 \\
 & -  3 g_{\Lambda}^4  \int_{\bd{q}}  [ {G}^4_{\Lambda} ( \bd{q} ) ]^{\bullet}.
 \label{eq:uLambdaflow}
 \end{align}
 \end{subequations}
Note that within our truncation
the flow equation  (\ref{eq:uLambdaflow}) for $u_{\Lambda}$ agrees with the
corresponding flow equation (\ref{eq:uLambdaflowm}) for constant $m$ because
we have neglected all vertices with more than four external legs.
For sharp momentum cutoff the flow equation (\ref{eq:flowfreeenergy})
for the field-independent part  $\tilde{\Gamma}^{(0)}_{\Lambda} =
{\Gamma}_{\Lambda} [ m_{\Lambda} ]$ of the Gibbs potential 
reduces to
\begin{equation}
 \partial_\Lambda \tilde{\Gamma}^{(0)}_{\Lambda} = 
 {\cal{V}} h \partial_{\Lambda} m_{\Lambda} 
 -  {\cal{V}}   \frac{  K_D }{2} \Lambda^{D-1} \ln \left( \frac{ c_0 \Lambda^2 + r_{\Lambda} }{ c_0 \Lambda^2 } \right),
  \label{eq:flowfreeenergy2}
  \end{equation}
while from Eq.~(\ref{eq:magflow}) we obtain for the flow of the 
magnetization
 \begin{equation}
 r_{\Lambda} \partial_{\Lambda} m_{\Lambda}  =  \frac{K_D}{2} \frac{ g_{\Lambda} \Lambda^{D-1}}{ c_0 \Lambda^2 + r_{\Lambda} }.
 \label{eq:mLambdaflow}
 \end{equation}
The flow of the other couplings is
 \begin{subequations}
 \label{eq:flowhcon2}
 \begin{align}
 \partial_{\Lambda} r_{\Lambda} = {} &  g_{\Lambda} \partial_{\Lambda} m_{\Lambda} - \frac{ K_D}{2} \frac{ u_{\Lambda} \Lambda^{D-1} }{ c_0 \Lambda^2 + r_{\Lambda} } 
  \nonumber
  \\
& + \frac{K_D}{2} \frac{ g_{\Lambda}^2 \Lambda^{D-1}}{ ( c_0 \Lambda^2 + r_{\Lambda} )^2 }, 
 \\
 \partial_{\Lambda} g_{\Lambda} = {} & u_{\Lambda} \partial_{\Lambda} m_{\Lambda}
 + \frac{ 3 K_D}{2}  \frac{ g_{\Lambda} u_{\Lambda} \Lambda^{D-1} }{ ( c_0 \Lambda^2 + r_{\Lambda} )^2 } 
 \nonumber
 \\
  & - K_D \frac{  g^3_{\Lambda} \Lambda^{D-1} }{ ( c_0 \Lambda^2 + r_{\Lambda} )^3 } ,
\\   
\partial_{\Lambda} u_{\Lambda} = {} & 
  \frac{ 3 K_D}{2}  \frac{ u^2_{\Lambda} \Lambda^{D-1} }{ ( c_0 \Lambda^2 + r_{\Lambda} )^2 } 
 - 6 K_D   \frac{ u_{\Lambda} g^2_{\Lambda} \Lambda^{D-1} }{ ( c_0 \Lambda^2 + r_{\Lambda} )^3 }
 \nonumber
 \\
  & + 3 K_D \frac{  g^4_{\Lambda} \Lambda^{D-1} }{ ( c_0 \Lambda^2 + r_{\Lambda} )^4 } .
  \end{align}
 \end{subequations}
The initial value $m_0 = m_0 ( h )$ for constant $h$ is given by the solution of the mean-field equation 
 \begin{equation}
  h = r_0 m_0 + \frac{ u_0 }{ 3!} m_0^3.
  \label{eq:hm0}
  \end{equation}
The initial values for the other vertices at constant $h$  are 
   \begin{subequations}
 \begin{align}
 \tilde{\Gamma}_{\Lambda_0}^{(0)} & = {\cal{V}} \left( \frac{{r}_0}{2} m_0^2 + \frac{ u_0}{ 4!} m_0^4 \right),
 \\
 \tilde{\Gamma}^{(2)}_{\Lambda_0} ( - \bd{k} , \bd{k} ) & = c_0 k^2 + r_{\Lambda_0} = c_0 k^2 + r_0 + \frac{u_0}{2} m_0^2,
 \\
 g_{\Lambda_0} & = u_0 m_0 , 
 \end{align} 
 \end{subequations}
where $m_0 = m_0 ( h )$ should be determined from the solution of the mean-field 
equation (\ref{eq:hm0}). 
 
\subsection{Scaling at the Ising fixed point for constant $h$}  
 
To find the fixed points and the scaling properties of the system 
we introduce the dimensionless rescaled magnetization
  \begin{align}
   m_l & = \sqrt{ \frac{ c_0}{K_D \Lambda^{D-2}}}  m_{\Lambda},
   \label{eq:mldef}
 \end{align}
and define  $r_l$, $g_l$ and $u_l$
as  in Eq.~(\ref{eq:couplingshscale}).
The rescaled magnetization $m_l$ 
replaces the  scale-dependent magnetic field $h_l$ 
of Sec.~\ref{sec:scalingh}   
as scale-dependent variable that quantifies the magnetism in the system for constant
external magnetic field. 
The rescaled form of the flow equations (\ref{eq:mLambdaflow}) and (\ref{eq:flowhcon2}) is then
 \begin{subequations}
 \label{eq:allflowh}
 \begin{align}
 \partial_l m_l & = \frac{ D-2 }{2} m_l - \frac{ g_l}{2 r_l ( 1 + r_l ) },
 \\
 \partial_l r_l & = 2 r_l + \frac{ u_l}{2 ( 1 + r_l ) }
 - \frac{ g_l^2}{ 2 (1 + r_l )} \left[ \frac{1}{ 1 + r_l } + \frac{1}{ r_l } \right],
 \\
 \partial_l g_l & = \frac{ 6-D}{2} g_l - \frac{ g_l u_l }{2 ( 1 + r_l ) }
  \left[ \frac{3}{ 1 + r_l } + \frac{1}{r_l } \right] + \frac{ g_l^3}{ ( 1 + r_l )^3 },  
 \\
 \partial_l u_l & = ( 4 - D ) u_l - \frac{ 3 u_l^2}{ 2 ( 1 + r_l )^2}
 +  \frac{ 6 u_l g_l^2}{ ( 1 + r_l )^3 } -  \frac{ 3 g_l^4}{ ( 1 + r_l )^4 }.
  \end{align} 
 \end{subequations}
For $m_l = g_l =0$ we recover again the one-loop flow equations of $\phi^4$ theory without magnetic field given in Eq.~(\ref{eq:ruflow}). As in Sec.~\ref{sec:RGm} we set
  $ v_l = g_l^2$ and linearize  the  flow around  of the Ising fixed point.
For small $\varepsilon$ we obtain up to corrections of order $\varepsilon^2$,
\begin{align}
 & \partial_l \begin{pmatrix} \delta r_l \\ \delta u_l \\ v_l \end{pmatrix} =
 \begin{pmatrix}  2 - \frac{\varepsilon}{3} & \frac{1}{2} + \frac{ \varepsilon}{12} 
  & \frac{3}{\varepsilon} + \frac{1}{2} -   \frac{\varepsilon}{12} \\
 0  & - \varepsilon & 4 \varepsilon \\
   0 & 0 & 6  - \varepsilon \end{pmatrix}  \begin{pmatrix} \delta r_l \\ \delta u_l \\ v_l \end{pmatrix} .
   \label{eq:flowlinh}
 \end{align}
Note that the third column of the matrix in Eq.~(\ref{eq:flowlinh}) differs from the
third column of the corresponding matrix 
for constant $m$ given in Eq.~(\ref{eq:flowlinm}).  
The eigenvalues of the matrix in Eq.~(\ref{eq:flowlinh}) are
 \begin{subequations}
 \begin{align}
 y_t & = 2 - \frac{ \varepsilon}{3} = \frac{ D+2}{3} ,
 \\
 y_s & = - \varepsilon,
 \\
 y_v & = 6 - \varepsilon = D +2 .
\end{align}
 \end{subequations}
The exponent  $y_v = D+2$ for constant $h$ differs
from the corresponding exponent $y_v = 2 - \varepsilon = D-2$ for constant $m$, see
Eq.~(\ref{eq:yvm}). The left-eigenvectors of the matrix (\ref{eq:flowlinh})
are to first order in $\varepsilon$ given by
 \begin{subequations}
\begin{align}
 \bd{\psi}_t^T &  = 
\left(  - \frac{  4 \varepsilon }{ 3 }, 
   - \frac{ \varepsilon }{ 3  }, \; 1   \right) ,
\\
 \bd{\psi}_s^T & = \left( 0 , 1 , - \frac{ 2 \varepsilon}{3} \right),
 \\
 \bd{\psi}_v^T & = ( 0,0,1 ).
\end{align}
 \end{subequations}
The corresponding scaling variables are
 \begin{subequations}
 \label{eq:tsvdef2}
  \begin{align}
 t_l & =  \bd{\psi}_t^T \begin{pmatrix} \delta r_l \\ \delta u_l \\ v_l \end{pmatrix} =  - \frac{4 \varepsilon}{3} \delta r_l  - \frac{ \varepsilon}{3}  \delta u_l 
 +  v_l ,
 \label{eq:tscale2}
 \\
 s_l & = \bd{\psi}_s^T \begin{pmatrix} \delta r_l \\ \delta u_l \\ v_l \end{pmatrix} = \delta u_l  - \frac{2 \varepsilon}{3}  v_l,
 \label{eq:sscale2}
 \\
 v_l & = \bd{\psi}_v^T \begin{pmatrix} \delta r_l \\ \delta u_l \\ v_l \end{pmatrix} .
 \end{align}
 \end{subequations}
While the relevant scaling variables $ t_l$ and $v_l$
agree with the corresponding scaling variables
for constant $m$ given in Eq.~(\ref{eq:tsvdef}), the irrelevant scaling variable
$s_l$ in Eq.~(\ref{eq:sscale2})
differs from the corresponding result $s_l = \delta u_l - 2 \varepsilon v_l$
obtained for constant $m$, see Eq.~(\ref{eq:sscale}).
The fact that for constant $m$ the exponent  $y_v$ has the value $D-2$ while for constant
$h$ its value is $D+2$ guarantees consistent results 
for the magnetic equation of state at the critical temperature. 
To see this,  we note that for constant
$h$ the coupling  $v_l = g_l^2$ is proportional to $h^2$ for small $h$. At the
critical temperature the
singular part of the Helmholtz free energy therefore satisfies the scaling law
 \begin{equation}
  F_{\rm sing}  ( h ) = b^{-D}  \Psi ( b^{y_v} h^2 ),
 \end{equation}
with some scaling function $\Psi ( x )$. With $b = h^{ - 2 / y_v}$  we obtain
 \begin{equation}
 F_{\rm sing} (h) = h^{2 D / y_v} \Psi (1 ) \propto h^{ 2 D / y_v }.
 \end{equation}
This gives the magnetic equation of state at the critical temperature in the form
 \begin{equation}
  m \propto \frac{ \partial F_{\rm sing} ( h ) }{\partial h } \propto h^{ \frac{ 2 D}{ y_v} -1}
   = h^{ 1/ \delta },
  \end{equation}
with
 \begin{equation}
\frac{1}{ \delta }=  \frac{2D}{ y_v} -1  = \frac{ 2 D - y_v }{y_v} = \frac{ D-2}{D+2},
 \end{equation}
in agreement with the result~(\ref{eq:deltares}) obtained for constant $m$.
Note that for constant $m$ the magnetic equation of  state at the critical temperature is directly generated in the form $ h ( m ) \propto m^{\delta}$, while  for constant $h$ we obtain
the inverse function  $ m ( h ) \propto h^{1/\delta}$.

\section{Phonon modified criticality: constant strain versus constant stress}
\label{sec:RGel}

In the previous sections \ref{sec:RGm} and \ref{sec:RGh} we have derived the
RG flow equations of $\phi^4$ theory for constant $m$ and for constant $h$. In spite of some
differences in the form of the flow equations, we have found that the linearized flow in the vicinity of the Ising fixed point is essentially identical, although the scaling  dimension of the field-induced three-point vertex is  different in both cases.
This reflects the fact that
for constant $m$ the critical magnetic equation of state is
obtained in the form $h \propto m^{\delta }$, whereas for constant $h$ the inverse function  $m \propto h^{1/ \delta}$ is generated.
We now consider another model system where the difference between the RG at constant order parameter and  constant conjugate  field is more drastic. The system is described by the following effective field theory involving an Ising order parameter $\phi (\bd{r} )$ which is coupled to another scalar field
$E ( \bd{r} )$ representing the volumetric strain of an isotropic elastic medium.   
In momentum space the bare action of our model can be written as \cite{Hansen25},
\begin{align}
	S[\phi,E] & =\frac{1}{2}\int_{\bd{k}} 
   \left[ 	
	( c_0 k^2 + r_0)  \phi_{-\bd{k}}\phi_{\bd{k}}+ F_0^{-1}(\bd{k})E_{-\bd{k}}E_{\bd{k}} \right]
	\nonumber 
	\\
	& +\frac{g_0}{2}\int_{\bd{k}_1}   \ldots  \int_{\bd{k}_3}{\cal{V}} \delta_{ \bd{k}_1 + \bd{k}_2 + \bd{k}_3 ,0 }\phi_{\bd{k}_1}\phi_{\bd{k}_2}E_{\bd{k}_3} \nonumber \\
	 & + \frac{u_0}{4!} \int_{\bd{k}_1}   \ldots  \int_{\bd{k}_4}
 {\cal{V}} \delta_{ \bd{k}_1 + \bd{k}_2 + \bd{k}_3 + \bd{k}_4 ,0 } \phi_{\bd{k}_1 } \phi_{\bd{k}_2} \phi_{\bd{k}_3}
 \phi_{\bd{k}_4},\label{eq:Sbare}
\end{align}
where 
the couplings $r_0$ and $u_0$ 
have already been introduced in Eq.~(\ref{eq:Sdef}), while $g_0$
parametrizes the strength of the coupling between the composite field $\phi^2$ and the strain field $E$. 
The inverse strain propagator is
\begin{align}
F_0^{-1}(\bd{k})=&\delta_{\bd{k},0}K_0+(1-\delta_{\bd{k},0})\rho_{0},\label{eq:free_E_propagator} 
 \end{align}
where $K_0$ is the bulk modulus and $\rho_0$ is the stiffness
for finite-momentum elastic fluctuations. For an isotropic elastic medium $\rho_0$ is 
larger than $K_0$ and is related to $K_0$ via 
 $\rho_0 = K_0 + \frac{2 (D-1)}{D} \mu$ where $\mu$ is the shear modulus \cite{Hansen25,Bruno80,Landau91,Chaikin95}.  
The volumetric strain $E ( \bd{r} ) = e + \epsilon ( \bd{r} )$  has a  uniform component  $e$ and a finite-momentum component  $\epsilon ( \bd{r} )$, so that
its  Fourier transform is  
\begin{equation}
	E_{\bd{k}}=   \int d^D r e^{ - i \bd{k} \cdot \bd{r} } E ( \bd{r} ) = \mathcal{V}e\delta_{\bd{k},0}+\left(1-\delta_{\bd{k},0}\right)\epsilon_{\bd{k}}.
\end{equation}
With this notation the quadratic strain part of the action (\ref{eq:Sbare})
can be written as
 \begin{align}
	& \frac{1}{2} \int_{\bd{k}} F_0^{-1} ( \bd{k}  ) E_{ - \bd{k} } E_{\bd{k}}   =\mathcal{V}\frac{K_0}{2}e^2+\frac{\rho_{0}}{2}\int_{\bd{k} \neq 0}\epsilon_{-\bd{k}}\epsilon_{\bd{k}}.
\end{align}

\subsection{FRG for constant strain}

\begin{figure}
 \begin{center}
  \centering
 \includegraphics[width=\columnwidth]{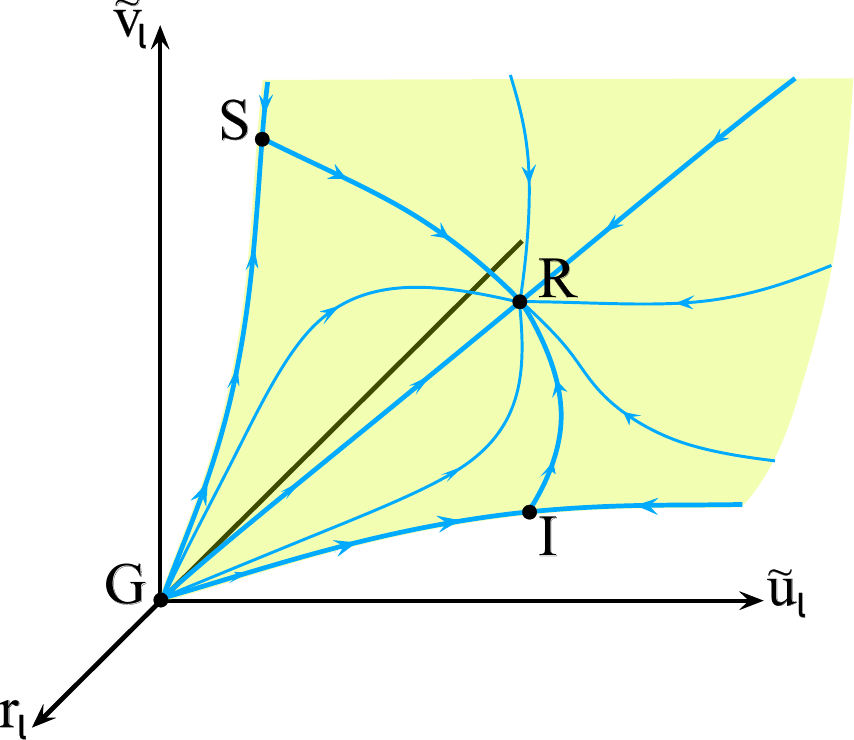}
   \end{center}
  \caption{%
Qualitative behavior of the RG flow for small $ \varepsilon = 4 - D > 0 $ at constant strain implied by Eqs.~(\ref{eq:flow_with_ueff}).
The flow has four distinct fixed points: Gaussian (G), Ising (I), spherical (S), and renormalized Ising (R).
Thick blue lines follow the eigendirections of the flow,
and the green sheet is the critical surface of R.
}
\label{fig:flowelasstrain}
\end{figure}

One-loop RG flow equations for the model (\ref{eq:Sbare}) for constant volumetric strain have been derived a long time ago by Bergman and Halperin \cite{Bergman76} and have recently been re-derived within the framework of the FRG in Ref.~[\onlinecite{Hansen25}]. Let us briefly recall the main results.
Close to $D = 4$ dimension it is sufficient to truncate the vertex expansion of the average effective action for constant strain $e$ as follows,
 \begin{widetext}
 \begin{align}
 \Gamma_{\Lambda} [ \bar{\phi} , e  + \bar{\epsilon}] \approx {} & \Gamma^{(0)}_{\Lambda}+\sigma_{\Lambda}\bar{\epsilon}_{\kv=0} + \frac{1}{2} \int_{\bd{k}}
 \left(  c_\Lambda k^2 + r_\Lambda  
\right)
 \bar{\phi}_{ - \bd{k}} \bar{\phi}_{\bd{k}} 
 + \frac{1}{2} \int_{\bd{k}} \left[ F_0^{-1} ( {\bd{k}} ) + \Pi_{\Lambda} ( \bd{k} ) \right]
 \bar{\epsilon}_{ - \bd{k}} \bar{\epsilon}_{\bd{k}} 
 \nonumber
 \\
 & +  \frac{g_{\Lambda}}{2} \int_{\bd{k}_1} \int_{\bd{k}_2} \int_{\bd{k}_3}
 {\cal{V}} \delta_{  \bd{k}_1 + \bd{k}_2 + \bd{k}_3 ,0}
  \bar{\phi}_{\bd{k}_1 }  \bar{\phi}_{\bd{k}_2} \bar{\epsilon}_{\bd{k}_3}
+   \frac{u_{\Lambda}}{4!} \int_{\bd{k}_1}  \int_{\bd{k}_2}   \int_{\bd{k}_3}   \int_{\bd{k}_4} {\cal{V}}
 \delta_{ \bd{k}_1 + \bd{k}_2 + \bd{k}_3 + \bd{k}_4 ,0}
 \bar{\phi}_{\bd{k}_1 } \bar{\phi}_{\bd{k}_2} \bar{\phi}_{\bd{k}_3} \bar{\phi}_{\bd{k}_4} .
 \label{eq:vertexp}
 \end{align}
 \end{widetext}
Here $\sigma_{\Lambda}$ is the renormalized pressure (i.e. the stress conjugate to the volumetric strain) $\Pi_{\Lambda} ( \bd{k} )$ is the scale-dependent elastic self-energy due to  fluctuations  of the Ising order parameter and the volumetric strain. 
The scale-dependence of the couplings
$r_{\Lambda}$, $g_{\Lambda}$ and $u_{\Lambda}$ is determined by RG flow equations given in
Ref.~[\onlinecite{Hansen25}].  Since we neglect the momentum-dependence of the self-energy of the Ising field the coupling $c_\Lambda = c_0$ is not renormalized.
It is then convenient to introduce the following dimensionless rescaled couplings~\cite{Hansen25}, 
\begin{subequations}
 \label{eq:couplings}
 \begin{align}
 r_l & = \frac{ r_\Lambda}{c_0 \Lambda^2 },
 \label{eq:rldef}
 \\
 g_l & = \sqrt{ \frac{ K_D}{ c_{0}^2 \rho_{\Lambda}  \Lambda^{4-D}} } g_{\Lambda},
 \label{eq:gldef}
  \\
 u_l & = \frac{K_D}{ c_0^2 \Lambda^{ 4-D}}u_{\Lambda},
 \label{eq:uldef}
 \\ 
 Y_{l} & = \frac{ \rho_0 }{ \rho_{\Lambda}  } = \frac{\rho_0}{  \rho_0 + \Pi_{\Lambda} ( 0 ) }
	\label{eq:Ydef}.
 \end{align} 
 \end{subequations}
where again
$l = \ln ( \Lambda_0 / \Lambda )$.  For constant strain these couplings satisfy the following system of
rescaled  flow equations \cite{Bergman76,Hansen25},
 \begin{subequations}
 \label{eq:flowruv}
 \begin{align}
 \partial_l r_l & = 2 r_l +  \frac{u_l-2g_l^2}{2 ( 1 + r_l )},
 \\
 \partial_l g_l & = \frac{ \varepsilon - y_l}{2} g_{l} - 
  \frac{\left( u_l  -  2 g_l^2\right) g_l}{2( 1 + r_l )^2 }  
  \nonumber
  \\
  & = \frac{\varepsilon }{2} g_l - 
  \frac{\left( 2 u_l  -  5 g_l^2\right) g_l}{4( 1 + r_l )^2 }  ,
  \label{eq:glflowmel} 
  \\
 \partial_l u_l & = \varepsilon u_l 
  - \frac{3} {2}\frac{\left(u_l^2 - 4 u_l  g_l^2 + 4 
  	g_l^4\right)}{ ( 1 + r_l )^2} ,
 \label{eq:ulflowmel}
\\
\partial_l Y_l & = - y_l Y_l,
 \end{align} 
 \end{subequations}
with
\begin{equation}
 y_l  = - \frac{ g_l^2}{2 ( 1 + r_l )^2 }.
 \label{eq:alphares}
 \end{equation}
In terms of
$ \tilde{v}_l  = g_l^2$ and
 $\tilde{u}_l  = u_l - 3 g^2_l = u_l - 3 \tilde{v}_l$,
the flow equations (\ref{eq:flowruv}) can alternatively be written as
\begin{subequations}
\label{eq:flow_with_ueff}
	\begin{align}
		\partial_l r_l & = 2 r_l +  \frac{\tilde{u}_l +\tilde{v}_l}{2 ( 1 + r_l )},
		\\
		\partial_l \tilde{u}_l & = \varepsilon \tilde{u}_l 
		- \frac{ 3 \tilde{u}_l^2}{ 2 ( 1 + r_l )^2} ,
		\label{eq:tildeuflow}
		\\
		\partial_l \tilde{v}_l & = \varepsilon \tilde{v}_l- \frac{\left(2\tilde{u}_l + \tilde{v}_l\right) \tilde{v}_l}{2( 1 + r_l )^2 } . 
		\label{eq:tildevflow}
	 \end{align} 
\end{subequations}
The RG flow implied by these flow equations has been discussed in Refs.~[\onlinecite{Bergman76,Hansen25}]. 
It turns out that for small $\varepsilon = 4-D > 0$ these flow equations have four fixed points:
apart from the Gaussian  fixed point G and the Ising fixed point I, the system exhibits two additional fixed points S (spherical) and R (renormalized Ising) where the coupling  $\tilde{v}_l$ approaches a finite limit
$\tilde{v}_{\ast} $ for $l \rightarrow \infty$. 
In Fig.~\ref{fig:flowelasstrain} we illustrate the RG flow
for $\varepsilon = 4 -D >0$.
In the following subsection we show that 
if we keep the stress (i.e. the pressure) constant during the RG transformation, then
only G and I remain accessible for physical initial conditions while
R and S move to an unphysical regime.

\subsection{FRG for constant stress}

If the stress $\sigma$ is fixed during the RG flow, the conjugate strain
$e_{\Lambda}$ must depend on the RG scale $\Lambda$.
To implement this boundary condition in the FRG, we follow the procedure outlined in Sec.~\ref{sec:RGh}. In analogy with Eq.~(\ref{eq:Gammashift})
we shift the strain field $\bar{E} ( \bd{r} ) = e_{\Lambda} + \bar{\epsilon} ( \bd{r} )$
and define the shifted average effective action
\begin{equation}
 \tilde{\Gamma}_{\Lambda} [ \bar{\phi} , \bar{\epsilon}  ] =
  {\Gamma}_\Lambda [ \bar{\phi} , e_{\Lambda} + \bar{\epsilon} ] 
  \label{eq:Gammashift2}
  \end{equation}
where the scale-dependent reference point $e_{\Lambda}$ should be chosen such that
 \begin{equation}   
   \left.  
 \frac{ \delta \tilde{\Gamma}_{\Lambda} [ \bar{\phi} , \bar{\epsilon} ]}{\delta \bar{\epsilon} (\bd{r})} \right|_{  \bar{\epsilon} ( \bd{r} )  = 0}  = \sigma,
  \label{eq:extremal3}
 \end{equation}
which is the analogue of Eq.~(\ref{eq:extremal2}).
As in Sec.~\ref{sec:RGh}, the scale-dependent reference point $e_{\Lambda}$ in the 
shifted average effective action (\ref{eq:Gammashift2}) generates an additional term in the flow equations involving the scale-derivative $\partial_{\Lambda} e_{\Lambda}$ of the strain.
In particular, from the condition that the one-point vertex is constant we obtain
 \begin{equation}
  \left[ K_0 + \Pi_{\Lambda} (0) \right] \partial_{\Lambda} e_{\Lambda} 
  = - \frac{g_{\Lambda}}{2} \int_{\bd{q}} \dot{G}_{\Lambda} ( \bd{q} ).
  \label{eq:strainflow}
  \end{equation}
which is the analogue of Eq.~(\ref{eq:magflow}). 
Note that the quantity $K_{\Lambda} = K_0 + \Pi_{\Lambda} (0 )$ on the left-hand side of Eq.~(\ref{eq:strainflow}) can be identified with the renormalized bulk modulus.
Moreover, the scale-dependence of $e_{\Lambda}$ modifies the flow of the coupling $r_{\Lambda}$ as follows
 \begin{equation}
 \partial_{\Lambda} r_{\Lambda} = g_{\Lambda} \partial_{\Lambda} e_{\Lambda}
 + \frac{u_{\Lambda}}{2} \int_{\bd{q}} \dot{G}_{\Lambda} ( \bd{q} )
 - g_{\Lambda}^2 \int_{\bd{q}} [ G_{\Lambda} ( \bd{q} ) F_{\Lambda} ( \bd{q} ) 
 ]^{\bullet},
 \label{eq:Sigmaflownew}
 \end{equation}
where  $F_{\Lambda} ( \bd{q} ) $ is the regularized strain propagator.
Within our truncation the flow equations for the mixed three-point vertex $g_{\Lambda}$ and the four-point vertex $u_{\Lambda}$ are not modified by the flow of the uniform strain so that  flow equations for the corresponding rescaled couplings $g_l$ and $u_l$ (as well as for $\tilde{u}_l$ and $\tilde{v}_l$) are the same as for fixed strain given
in Eqs.~(\ref{eq:glflowmel}) and (\ref{eq:ulflowmel}).
From Eq.~(\ref{eq:strainflow}) we find that the rescaled uniform strain
 \begin{equation}
 e_l = \sqrt{ \frac{ \rho_\Lambda}{ K_D \Lambda^D}} e_{\Lambda}
\end{equation}
satisfies the flow equation
 \begin{equation}
 \partial_l e_l = \frac{ D + y_l}{2} e_l - \frac{ X_l g_l}{2 ( 1 + r_l ) },
 \end{equation} 
where the dimensionless coupling
 \begin{align}
  X_l  & = \frac{ \rho_{\Lambda}}{K_{\Lambda}} = 
  \frac{\rho_{0} + \Pi_{\Lambda} ( 0 ) }{K_0 + \Pi_{\Lambda} (0) }
  \nonumber
  \\
  &  = \frac{\rho_{\Lambda}}{ \rho_{\Lambda} - \frac{2 (D-1)}{D} \mu } 
   = \frac{1}{ 1 - \frac{ 2 (D-1)}{D} \frac{\mu}{\rho_0} Y_{l}}   
  \end{align} 
satisfies the Ricatti equation
 \begin{equation}
 \partial_l X_l = y_l X_l ( 1 - X_l ) ,
 \label{eq:Ricatti}
 \end{equation}
where $ y_l $ is given by Eq.~(\ref{eq:alphares}) as for constant strain. 
Moreover, using Eqs.~(\ref{eq:strainflow}) and (\ref{eq:Sigmaflownew})
we find that for constant stress the coupling $r_l$ satisfies
 \begin{equation}
 \partial_l r_l = 2 r_l + \frac{ u_l - (2 + X_l) g_l^2  }{2 ( 1 + r_l )}
 = 2 r_l + \frac{ \tilde{u}_l + (1 - X_l) \tilde{v}_l  }{2 ( 1 + r_l )},
 \label{eq:rpressure}
 \end{equation}  
Note that in contrast to the flow at constant strain,
at constant stress the renormalized elastic stiffness $ \rho_\Lambda = \rho_0 / Y_l $ couples to the flow of the remaining coupling constants.  
In the following,
we show that this has a drastic effect on the fixed point structure of the theory.
\begin{table}%
\begin{tabular}{c|c|c|c|c|c}%
fixed point 
&
$ r_* $
&
$ \tilde{u}_* $
&
$ \tilde{v}_* $ 
&
$ X_* $
&
$ y_* $
\\
\hline
\hline
G (Gaussian) & 0 & 0 & 0 & $ X_0 $ & 0
\\
\hline
I (Ising) 
& 
$ \;\; - \varepsilon / 6 \;\; $
&
$ \;\;  2 \varepsilon / 3 \;\; $
&
$ \;\; 0 \;\; $
&
$ \;\; X_0 \;\; $
&
0
\\
\hline
R (renorm.~Ising)
&
$ - \varepsilon / 3 $
&
$ 2 \varepsilon / 3 $
&
$ 2 \varepsilon / 3 $
&
$  \;\; 0  \;\; $
&
$ \;\; - \varepsilon / 3 \;\; $
\\
\hline
$ \tilde{\text{R}} $ (renorm.~Ising)
&
$ \;\; - \varepsilon / 6 \;\; $
&
$ \;\; 2 \varepsilon / 3 \;\; $
&
$ \;\; 2 \varepsilon / 3 \;\; $
&
$ \;\; 1 \;\; $
&
$ \;\; - \varepsilon / 3 \;\; $
\\
\hline
S (spherical)
&
$ \;\; - \varepsilon / 2 \;\; $
&
$ \;\; 0 \;\; $
&
$ \;\; 2 \varepsilon \;\; $
&
0
&
$ \;\; - \varepsilon \;\; $
\\
\hline
$ \tilde{\text{S}} $ (spherical)
&
$ 0 $
&
$ 0 $
&
$ 2 \varepsilon $
&
1
&
$ - \varepsilon $
\\
\hline
\end{tabular}%
\caption{Fixed point values of the rescaled couplings of the RG flow at constant stress implied by 
Eqs.~(\ref{eq:tildeuflow}, \ref{eq:tildevflow}, \ref{eq:Ricatti} and \ref{eq:rpressure}),
to first order in 
$ \varepsilon = 4 - D $.
Compared to the flow at constant strain sketched in Fig.~\ref{fig:flowelasstrain}, Gaussian and Ising fixed points broaden into a manifold labeled by the the initial value $ X_0 $ of the rescaled coupling $ X_l $ associated with the elastic stiffness.
Furthermore, the renormalized Ising and pherical fixed points are split into two distinct fixed points for different values of $ X_* $,
denoted R and $ \tilde{\text{R}} $
and S and $ \tilde{\text{S}} $,
respectively. 
Only the Gaussian and Ising fixed points can be reached from physical initial conditions with $  X_0 > 1 $,
rendering the renormalized Ising and spherical fixed points that are accessible for constant strain unphysical for constant stress;
see Fig.~\ref{fig:flow_X}.
}
\label{table:fixed_points}
\end{table}%

Because for both fixed points  G and I the value of
$ \tilde{v}_*$ and hence also $y_* $ vanishes, the flow equation~(\ref{eq:Ricatti})
for
$ X_l $ implies that $X_l =X_0$ remains fixed to its initial value $ X_0 $. 
Thus,
G and I are actually fixed point manifolds.
As a result,
the thermal scaling variable close to I depends to first order in $ \varepsilon $ on the position $ X_0 $ on this manifold as
\begin{equation}%
t_l = 
\left( 4 - \varepsilon \right) \delta r_l
+
\left( 1 - \frac{ 2 \varepsilon }{ 3 } \right) \delta \tilde{u}_l
+
\left( 1 - X_0 \right) \delta \tilde{v}_l .
\end{equation}%
For $ \varepsilon = 0 $,
this expression also gives the thermal scaling variable close to G.
It reduces to the constant-strain result of Ref.~\cite{Hansen25} only for the unphysical case of vanishing elastic stiffness, 
$ X_0 = \rho_0 / K_0 = 0 $.

\begin{figure}
 \begin{center}
  \centering
 \includegraphics[width=\columnwidth]{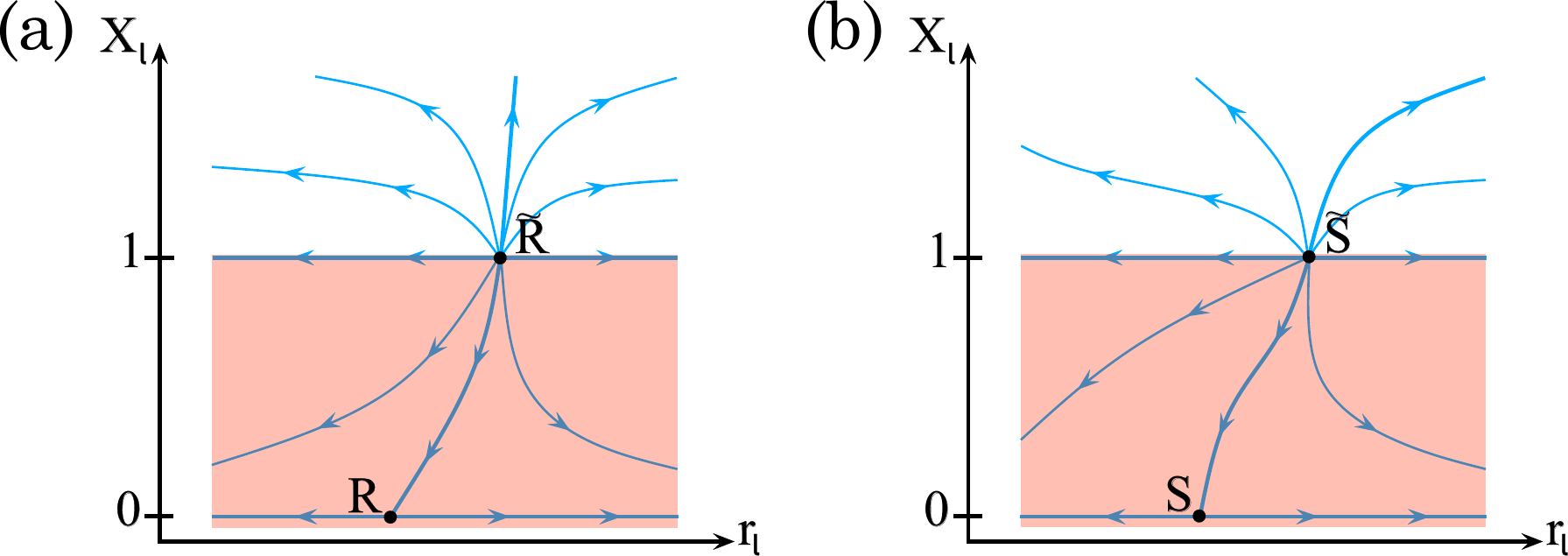}
   \end{center}
  \caption{%
Sketch of the RG flow for constant stress implied by 
Eqs.~(\ref{eq:tildeuflow}, \ref{eq:tildevflow}, \ref{eq:Ricatti} and \ref{eq:rpressure})
close to 
(a) the renormalized Ising and
(b) the spherical fixed points,
in the relevant $ r_l $-$ X_l $ plane.
The other couplings $ \tilde{u}_l $ and $ \tilde{v}_l $ are set to their respective fixed point values;
see Table~\ref{table:fixed_points}.
The red region with $ X_l \le 1 $ is physically forbidden by thermodynamic stability~\cite{Landau91},
and inaccessible due to the repulsive fixed points $ \tilde{\text{R}} $ and $ \tilde{\text{S}} $,
which generate a runaway flow towards $ X_l \to \infty $ in the physically allowed regime.
}
\label{fig:flow_X}
\end{figure}

Next,
let us investigate the fate of the renormalized Ising and spherical fixed points.
It is obvious from the flow equation \eqref{eq:Ricatti} 
that any fixed point with $ \tilde{v}_* \neq 0 $ and hence $ y_* \neq 0 $
must have $ X_* = 0 $ or $ 1 $.
In the former case,
the fixed-point equations are formally identical to the equations for constant strain discussed in detail in Ref.~\cite{Hansen25}.
Therefore,
both R and S still exist,
but are located at the unphysical value 
$ X_* = 0 $.
Analog solutions can also be obtained for the second possible value $ X_* = 1 $.
We will denote these by $ \tilde{\text{R}} $ and $ \tilde{\text{S}} $,
respectively,
because to first order in $ \varepsilon $ they retain the same fixed point values 
$ r_* $, $ \tilde{u}_* $, $ \tilde{v}_* $, and $ y_* $ 
of the other couplings. 
This indicates that the constant-strain fixed points R and S actually split into two distinct fixed points at constant stress;
see Table~\ref{table:fixed_points} for a summary of all fixed points of the constant-stress flow equations (\ref{eq:tildeuflow}, \ref{eq:tildevflow}, \ref{eq:Ricatti} and \ref{eq:rpressure}).
Note in particular that the four fixed points R, $ \tilde{\text{R}} $, S, and $ \tilde{\text{S}} $ with $ \tilde{v}_* $, $ y_* \neq 0 $ lie outside the thermodynamically stable regime 
$ X_0 = \rho_0 / K_0 > 1 $~\cite{Landau91}. 
To determine what this means physically,
we now consider the linearized flow around these four fixed points.
For R,
we obtain
\begin{equation}%
\partial_l
\begin{pmatrix}%
\delta r_l \\
\delta \tilde{u}_l \\
\delta \tilde{v}_l \\
\delta X_l 
\end{pmatrix}%
=
\begin{pmatrix}%
2 - \frac{ 2 \varepsilon }{ 3 }
&
\frac{1}{6} + \frac{ \varepsilon }{ 6 }
&
\frac{1}{6} + \frac{ \varepsilon }{ 6 }
&
- \frac{ \varepsilon }{ 3 }
\\
0 & - \varepsilon & 0 & 0
\\
0 
& 
- \frac{ 2 \varepsilon }{ 3 }
& 
- \frac{ \varepsilon }{ 3 }
&
0
\\
0 & 0 & 0 &  - \frac{ \varepsilon }{ 3 }
\end{pmatrix}%
\begin{pmatrix}%
\delta r_l \\
\delta \tilde{u}_l \\
\delta \tilde{v}_l \\
\delta X_l 
\end{pmatrix}%
,
\end{equation}%
which has only one relevant eigenvalue associated with the thermal scaling variable $ t_l $.
For $ \tilde{\text{R}} $ on the other hand,
the linearized flow is given by
\begin{equation}%
\partial_l
\begin{pmatrix}%
\delta r_l \\
\delta \tilde{u}_l \\
\delta \tilde{v}_l \\
\delta X_l 
\end{pmatrix}%
=
\begin{pmatrix}%
2 - \frac{ \varepsilon }{ 3 }
&
\frac{1}{2} + \frac{ \varepsilon }{ 12 }
&
0
&
- \frac{ \varepsilon }{ 3 }
\\
0 & - \varepsilon & 0 & 0
\\
0 
& 
- \frac{ 2 \varepsilon }{ 3 }
& 
- \frac{ \varepsilon }{ 3 }
&
0
\\
0 & 0 & 0 &  \frac{ \varepsilon }{ 3 }
\end{pmatrix}%
\begin{pmatrix}%
\delta r_l \\
\delta \tilde{u}_l \\
\delta \tilde{v}_l \\
\delta X_l 
\end{pmatrix}%
.
\end{equation}%
This shows that for $ \tilde{\text{R}} $,
not only the thermal scaling variable $ t_l $ but also $ X_l $ is a relevant direction. 
Put together,
this implies a flow from $ \tilde{\text{R}} $ at $ X_* = 1 $ toward R at $ X_* = 0 $,
sketched in Fig.~\ref{fig:flow_X} (a) in the relevant $ r_l $-$ X_l $ plane.

A similar picture emerges for S and $ \tilde{\text{S}} $,
where the linearized flow equations are
\begin{equation}%
\partial_l
\begin{pmatrix}%
\delta r_l \\
\delta \tilde{u}_l \\
\delta \tilde{v}_l \\
\delta X_l 
\end{pmatrix}%
=
\begin{pmatrix}%
2 - \varepsilon
&
\frac{1}{2} + \frac{ \varepsilon }{ 4 }
&
\frac{1}{2} + \frac{ \varepsilon }{ 4 }
&
- \varepsilon
\\
0 & \varepsilon & 0 & 0
\\
0 
& 
- 2 \varepsilon 
& 
- \varepsilon
&
0
\\
0 & 0 & 0 &  - \varepsilon 
\end{pmatrix}%
\begin{pmatrix}%
\delta r_l \\
\delta \tilde{u}_l \\
\delta \tilde{v}_l \\
\delta X_l 
\end{pmatrix}%
,
\end{equation}%
and
\begin{equation}%
\partial_l
\begin{pmatrix}%
\delta r_l \\
\delta \tilde{u}_l \\
\delta \tilde{v}_l \\
\delta X_l 
\end{pmatrix}%
=
\begin{pmatrix}%
2 
&
\frac{1}{2} 
&
0
&
- 2 \varepsilon
\\
0 & \varepsilon & 0 & 0
\\
0 
& 
- 2 \varepsilon 
& 
- \varepsilon
&
0
\\
0 & 0 & 0 &  \varepsilon 
\end{pmatrix}%
\begin{pmatrix}%
\delta r_l \\
\delta \tilde{u}_l \\
\delta \tilde{v}_l \\
\delta X_l 
\end{pmatrix},
\end{equation}%
respectively.
Again,
the flow is repelled from $ \tilde{\text{S}} $ at $ X_* = 1 $ and attracted toward S at $ X_* = 0 $;
see Fig.~\ref{fig:flow_X} (b).
The main difference of the renormalized Ising and the spherical fixed points is the existence of an additional relevant scaling variable in the $ \tilde{u}_l $-$ \tilde{v}_l $ plane for the latter,
similar to the constant-strain case sketched in Fig.~\ref{fig:flowelasstrain}.

It is now important to keep in mind that thermodynamic stability requires that the initial value of the rescaled coupling $ X_l $ associated with the elastic stiffness satisfies $ X_0 > 1 $ for real systems~\cite{Landau91}.
Therefore,
they will always be repelled from $ \tilde{\text{R} } $ and $ \tilde{\text{S} } $ toward larger values of $ X_l $,
resulting in a runaway flow.
Thus, a thermodynamically stable physical system can never enter the basin of attraction of the R and S fixed points,
and the whole unphysical $ X_l \le 1 $ region is shielded by the existence of the repulsive $ \tilde{\text{R} } $ and $ \tilde{\text{S} } $ fixed points.
This is qualitatively shown in Fig.~\ref{fig:flow_X}.
The runaway flow can also be understood analytically from the flow equation~(\ref{eq:Ricatti}).
Its formal solution is given by
\begin{equation}%
\label{eq:Xsol}
X_l = 
\frac{ 
X_0 e^{ \int_0^l dt y_t } 
}{
1 + X_0 \left( e^{ \int_0^l dt y_t } - 1 \right) 
} .
\end{equation}%
In the vicinity of the renormalized Ising and spherical fixed points,
$ y_l $ approaches
a finite limit $ y_* <0 $ for $l \rightarrow \infty$;
see Table~\ref{table:fixed_points}.
Then we may estimate the integrals 
in Eq.~(\ref{eq:Xsol}) 
by approximating $y_l \approx y_*$.
However,
keeping in mind that
$ y_* < 0 $ and $ X_0 > 1 $,
we conclude that $ X_l $ diverges at a finite scale $l_*$ given by 
\begin{equation}%
l_* = \frac{1}{ \left| y_* \right| } 
\ln \left( \frac{ X_0 }{ X_0 - 1 } \right) .
\end{equation}%
At this scale the RG exhibits a runaway flow for any physical initial condition $ X_0 > 1 $.

We conclude that
the fixed points S and R can be realized only for constant strain. 
If we impose the boundary condition of constant stress,
they can never be reached from physical initial conditions,
and instead cause a runaway flow.
This is consistent with the classic analysis of Bergman and Halperin~\cite{Bergman76}, and by Bruno and Sak~\cite{Bruno80}, 
as well as earlier work by Larkin and Pikin \cite{Larkin69} 
who found that for constant pressure (which is the stress conjugate to the volumetric strain) the coupling of an Ising order parameter to elasticity gives rise to a first order transition signaled by a runaway flow.  
The physical origin for the qualitative difference between the RG flows at constant strain and constant stress is that in the latter case the fluctuations of the macroscopic strain give rise to additional long-range interactions which tend to favor first-order transitions \cite{Larkin69,Bergman76,Sarkar23}. Within our FRG approach the extra interaction between Ising fluctuations due to fluctuations of the macroscopic strain can be derived by solving Eq.~(\ref{eq:strainflow}) for
$\partial_{\Lambda} e_{\Lambda}$ and substituting the result into the flow equation (\ref{eq:Sigmaflownew}) to obtain an effective flow equation for $r_{\Lambda}$ for constant stress,
 \begin{equation}
 \partial_{\Lambda} r_{\Lambda} = 
  \frac{1}{2} \left(
 u_{\Lambda} -   \frac {g_\Lambda^2}{K_{\Lambda} } \right) \int_{\bd{q}} \dot{G}_{\Lambda} ( \bd{q} )
 - g_{\Lambda}^2 \int_{\bd{q}} [ G_{\Lambda} ( \bd{q} ) F_{\Lambda} ( \bd{q} ) 
 ]^{\bullet}.
 \label{eq:Sigmaflownewtoo}
 \end{equation}
Hence, fluctuations of the uniform strain effectively
replace the interaction parameter $u_{\Lambda}$ by $u_{\Lambda} - g^2_{\Lambda} / K_{\Lambda}$. The runaway flow in terms of rescaled couplings is due to the vanishing 
of the renormalized bulk modulus $K_{\Lambda}$ at a finite scale.

\section{Summary and conclusions}
\label{sec:summary}

RG transformations of systems subject to finite external fields can be performed either by fixing the field during the RG transformation, or by fixing the
conjugate order parameter. In this work we have shown how to implement these distinct boundary conditions  within the framework of the FRG. 
To illustrate the main ideas, we have considered the RG flow of $\phi^4$ theory in a magnetic field $h$ slightly below four dimensions, where the Ising fixed point is accessible perturbatively. Although the flow equations for constant $h$ and for constant magnetization  $m$ do not agree, and in spite of the fact that the value of the scaling exponent
$y_v$ 
associated with the linearized flow of the three-point vertex close to the Ising fixed point changes when we fix $h$ instead of $m$,
the final result for the critical exponent $\delta$ agrees in both cases. 

However, in general the fixed point structure of a given model can qualitatively change when the external field instead of the order parameter is held constant.  To illustrate this, we have considered
a $\phi^4$ theory coupled to elasticity. In this case either the strain or the conjugate stress can be held constant  during the RG transformation. While for constant strain one obtains four distinct fixed points close to four dimensions, for constant stress only the trivial Gaussian fixed point and the usual Ising fixed point survive for real systems,
but broaden into a manifold.
The other two fixed points move to a physically inaccessible parameter regime and generate a runaway flow that describes a first-order phase transition~\cite{Bergman76, Bruno80}. 
This shows that the existence of critical points and the nature of phase transitions can crucially depend on the boundary conditions. 

Another class of systems  where our results are likely to be relevant are interacting electrons in condensed matter \cite{Volovik19}. 
Usually, many-body perturbation theory for interacting electrons is formulated at fixed density, which implies that the chemical
potential has to be adjusted to keep the density constant  when the RG transformation is carried out. The FRG formalism  developed in Sec.~\ref{sec:RGm} and Sec.~\ref{sec:RGh} can  be generalized  to derive  formally exact FRG flow equations for interacting fermions at constant density and constant chemical potential. Especially for extremely strong
 interactions  we expect significant differences in the RG flows \cite{Arnold25a,Arnold25b}.

\section*{ACKNOWLEDGEMENTS}
	
We acknowledge illuminating discussions with Markus Garst and thank the 
Deutsche Forschungsgemeinschaft (DFG, German Research Foundation) for financial support via  TRR 288 - 422213477.


\begin{thebibliography}{99}
\bibitem{footnotefunc}
In momentum space we normalize the functional derivative symbol $\frac{\delta}{\delta \phi_{\bd{k}} }$ as follows: 
$\frac{\delta}{\delta \phi_{\bd{k}} } = {\cal{V}} \frac{ \partial }{\partial \phi_{\bd{k}}}$,
where $\frac{\partial}{\partial \phi_{\bd{k}}}$ is the usual partial derivative.
The factor of volume is convenient because then
$\frac{\delta}{\delta \phi_{\bd{k}} }  \phi_{ \bd{k}^{\prime} } =
{\cal{V}} \frac{\partial}{ \partial \phi_{\bd{k}} }   \phi_{\bd{k}^{\prime}}  
= {\cal{V}} \delta_{  \bd{k} , \bd{k}^\prime}  \rightarrow  ( 2 \pi )^D \delta ( \bd{k} - \bd{k}^\prime ) $ for ${\cal{V}} \rightarrow \infty$.
\bibitem{remarkfunctional}
Note that in Ref.~[\onlinecite{Schuetz06}]
we have  used a slightly different definition of the
the generating functional of irreducible vertices for constant magnetic field, which amounts to
subtracting the contribution from the one-point vertex, so that the
one-point vertex vanishes identically at all scales. The resulting flow equations for the
order parameter and all higher-order vertices agree with the flow equations
of Sec.~\ref{sec:RGh}.
\end{thebibliography}
\end{document}